\documentclass[aps,prd,preprint,superscriptaddress,floatfix,nofootinbib]{revtex4}

\usepackage[dvips]{graphicx}

\newcommand{\lsim}{\mathrel{\mathop{\kern 0pt \rlap
  {\raise.2ex\hbox{$<$}}}
  \lower.9ex\hbox{\kern-.190em $\sim$}}}
\newcommand{\gsim}{\mathrel{\mathop{\kern 0pt \rlap
  {\raise.2ex\hbox{$>$}}}
  \lower.9ex\hbox{\kern-.190em $\sim$}}}

\begin{document}

\preprint{DFTT 06/2002}
\preprint{ROM2F/2002/02}

\title{~\\ Effect of the galactic halo modeling on the DAMA/NaI annual 
 modulation result: an extended analysis of the data for  
WIMPs with a purely spin-independent coupling}


  
\author{P. Belli}
\email{belli@roma2.infn.it}
\affiliation{Dipartimento di Fisica, Universit\`a di Roma ``Tor Vergata''
    and INFN, Sezione di Roma2, I--00133 Roma, Italy} 

\author{R. Cerulli}
\email{cerulli@roma2.infn.it}
\affiliation{Dipartimento di Fisica, Universit\`a di Roma ``Tor Vergata''
    and INFN, Sezione di Roma2, I--00133 Roma, Italy}

\author{N. Fornengo} 
\email{fornengo@to.infn.it}
\homepage{http://www.to.infn.it/~fornengo/}
\affiliation{Dipartimento di Fisica Teorica, Universit\`a di Torino 
and INFN, Sezione di Torino, Via P. Giuria 1, I--10125 Torino, Italy}

\author{S. Scopel} 
\email{scopel@to.infn.it}
\affiliation{Dipartimento di Fisica Teorica, Universit\`a di Torino 
and INFN, Sezione di Torino, Via P. Giuria 1, I--10125 Torino, Italy}

\date{\today}

\begin{abstract} \vspace{1cm}  
  The DAMA/NaI Collaboration has observed a 4--$\sigma$ C.L. model independent
  effect investigating the annual modulation signature in the counting rate of
  an NaI(Tl) set--up (total exposure of 57986 kg day) and the implications of
  this effect have been studied under different model--dependent assumptions.
  In this paper we extend one of the previous analyses, the case of a WIMP with
  a purely spin-independent coupling, by discussing in detail the implications
  on the results of the uncertainties on the dark matter galactic velocity
  distribution.  We study in a systematic way possible departures from the
  isothermal sphere model, which is the parameterization usually adopted to
  describe the halo. We specifically consider modifications arising from
  various matter density profiles, effects due to anisotropies of the velocity
  dispersion tensor and rotation of the galactic halo. The hypothesis of WIMP
  annual modulation, already favoured in the previous analysis using an
  isothermal sphere, is confirmed in all the investigated scenarios, and the
  effects of the different halo models on the determination of the allowed
  maximum--likelihood region in the WIMP mass and WIMP--nucleon cross--section
  are derived and discussed.
\end{abstract}


\maketitle

\section{Introduction}
\label{sec:intro}

In the last few years various technical approaches have been exploited
in direct searches for Weakly Interacting Massive Particles (WIMP)
\cite{direct_exp}. In particular, the DAMA/NaI Collaboration has
collected a very large body of statistics which allows one to look for
the distinctive signature in Dark Matter (DM) direct detection
represented by the annual modulation of the rate
\cite{damalast,syst,noi_belli}, an effect which is due to the rotation
of the Earth around the Sun \cite{freese}.  The analysis of the
DAMA/NaI data after 4 years of running \cite{damalast,noi_belli},
corresponding to a total exposure of 57986 kg day, has indeed led to
the observation of an annual--modulation effect, which does not appear
to be related to any possible source of systematics \cite{syst}. This
exciting result has been analyzed under different hypotheses on the
properties of WIMP dark matter: purely spin independent coupling
\cite{damalast}, mixed spin--coherent interaction \cite{dama_mixed},
inelastic dark matter \cite{dama_inelastic}. In the case of purely
spin--independent interactions, the annual modulation result has been
shown to be compatible with a galactic halo composed, at least
partially, by relic neutralinos in different classes of supersymmetric
models \cite{noi,altri}.

In the present analysis we consider the case of a WIMP with coherent 
interactions dominant over the spin--dependent ones.
The analysis of the counting rate of any direct detection experiment may be
done in terms of the WIMP mass $m_{W}$ and of the quantity $\xi
\sigma_{\rm scalar}^{\rm (nucleon)}$, where $\sigma_{\rm scalar}^{\rm (nucleon)}$ is the
WIMP--nucleon cross section for scalar interaction and $\xi$
\footnote{In Ref. \cite{damalast} the same symbol indicates a
different quantity: $\xi=\rho_W/({\rm 0.3\; GeV cm^{-3}})$.} is the
fractional amount of local non--baryonic DM density which is ascribed
to the WIMP responsible for the effect ($\xi \leq 1$)
\cite{damalast,noi}.  Performing a maximum--likelihood analysis of the
data, the DAMA/NaI Collaboration has derived a region in the plane
$m_{W}$--$\xi \sigma_{\rm scalar}^{\rm (nucleon)}$ which is compatible
at 3--$\sigma$ C.L. to the observed effect of annual modulation. The
properties of this region are sensitive to astrophysical inputs
\cite{noi_belli,depart1}, some of which deserve a reanalysis and a
deeper insight.

One of the main ingredients for the calculation of the expected rates is the
distribution function of WIMPs in their six--dimensional phase space: $F(\vec r, \vec
v)\;d^3r\;d^3v$ (where the position vector $\vec r \equiv (x,y,z)$ and the
velocity vector $\vec v \equiv (v_x,v_y,v_z)$ are defined in the rest frame of
the Galaxy). Direct detection rates $R_{det}$ depend on the distribution function
(DF) at the Earth position in the Galaxy:
\begin{equation}
f(\vec v) \equiv F(\vec R_0,\vec v)\, , 
\label{eq:df}
\end{equation}
where $\vec R_0 \equiv (R_0,0,0)$ is the location of the Earth at a distance
$R_0 \simeq 8.5$ kpc from the galactic center and along the galactic plane. It
is therefore clear that an accurate calculation of the expected detection rates
requires a knowledge of the phase--space distribution function $F(\vec r, \vec
v)\;d^3r\;d^3v$.

From the observational side, the most relevant piece of information coming from
astrophysics is related to the rotational velocity of objects bounded to the
Galaxy:
\begin{equation}
v_{rot}^2(r) = \frac{G M_{tot}(r)}{r},
\label{eq:vrot}
\end{equation}
where $G$ is the Newton's constant and $M_{tot}(r)$ denotes the total mass contained
inside the radius $r \equiv |\vec r|$:
\begin{equation}
M_{tot}(r) = \int_{r^{\prime}<r}d^3 r^{\prime} \rho_{tot}(\vec{r^{\prime}}).
\end{equation}
The rotational velocity therefore depends on the total matter density
distribution in space:
\begin{equation}
\rho_{tot}(\vec r) = \rho_{DM}(\vec r)+\rho_{vis}(\vec r),
\label{eq:rho_total}
\end{equation} 
where the DM density distribution function is:
\begin{equation}\rho_{DM}(\vec r) \equiv \int d^3v \;F(\vec r, \vec v),
\label{eq:rho_dm}
\end{equation} 
and $\rho_{vis}(\vec r)$ represents the contribution to the matter
density due to components other than the DM, like the disk and the
bulge. The local values for the rotational velocity and for the DM
matter density are denoted by $v_0 \equiv v_{rot}(\vec R_0)$ and
$\rho_0\equiv\rho_{DM}(\vec R_0)$ and they represent two key
parameters in the calculation of WIMP direct detection rates, as it
will be discussed in the following.

In order to calculate the DF of Eq. (\ref{eq:df}) one must invert
Eq. (\ref{eq:rho_dm}) taking into account observational data.
This problem is affected by degeneracies
that cannot be solved without adding some
piece of information. This explains why the velocity distribution represents one of the
main source of uncertainty in the calculation of direct detection signals.

The usual approach to this problem consists in assuming that the
system has some symmetry and that the distribution $F$ depends on the
phase space parameters only through some integrals of motion
(energy, angular momentum): this last condition automatically implies
stationarity and that the Jeans' equations are verified \cite{binney}.
The velocity ellipsoid \mbox{$\sigma_{ij}\equiv <v_i v_j>$} may than
be calculated as a function of the derivatives of the potential (which
are related to the rotational velocity) by making use of Euler's
equation \cite{binney}. Physically, this corresponds to impose
hydrostatic equilibrium between pressure and gravitational attraction.
 
The most common and widely used example of such a procedure, and by far the
simplest, is the isothermal sphere model. It consists in a spherical infinite
system with a flat rotational curve, which automatically implies $\rho(r)
\propto r^{-2}$ and the potential $\Psi \propto \ln(r^2)$ .  The DF
may be easily worked out, and turns out to be a Maxwellian: $f(v) \propto
\exp(- 3 v^2/(2 v^2_{rms}))$, where $v\equiv |\vec v|$ and $v_{rms}$ denotes
the root mean squared velocity of the WIMPs. The isothermal sphere describes a
self--gravitating gas of collisionless particles in thermal equilibrium,
representing the highest entropy rearrangement of WIMPs in their phase space. A
strong argument in favour of this last property is the ``violent relaxation''
model of Lynden--Bell \cite{violent_relax}, which indicates that the violently
changing gravitational field of the newly formed Galaxy may have lead the
non--interacting WIMPs to thermal equilibrium. Hydrostatic equilibrium and the
assumption that the velocity ellipsoid is isotropic allows to calculate
$v_{rms}$ through the relation: $v_{rms}^2=3/2 \; v_{rot}^2(R_0)$.  Due to its
simplicity, the isothermal sphere model has become the ``standard'' assumption
in the evaluation of DM expected rates, and has been used extensively in the
literature, including the analysis of the DAMA/NaI modulation data
\cite{damalast}.  However many of its underlying assumptions (sphericity of the
halo, absence of rotation, isotropy of the dispersion tensor, flatness of the
rotational curve) are not strongly constrained by astrophysical observations.
Moreover the isothermal sphere is strictly unphysical and can only represent
the behaviour of the inner part of physical systems, since it has a total
infinite mass and needs some cut--off at large radii.

In light of the latest experimental data on WIMP direct searches, the
issue of possible departures from the isothermal sphere model has
gained interest and prompted several discussions
\cite{noi_belli,depart1,anis}.  In the present paper we intend to
analyze this issue in a systematic way, by employing a comprehensive
set of self--consistent galactic halo models.  Each model introduces a
different degree of deviation from the simple isothermal sphere. We
consider modifications in the velocity distribution function which are
originated from a change of the gravitational potential or a change of
the DM density profile
\cite{evans_pl,evans_log,nfw,moore,kravtsov,jaffe,triaxial}. We
classify the different models depending on the symmetry properties of
the galactic halo: spherical potential/density profile with an
isotropic velocity dispersion; spherical potential/density profile
with a non--isotropic velocity dispersion; axisymmetric models;
triaxial models. For the axisymmetric models we also consider the
possibility of having a co--rotating or counter--rotating halo.

The plan of the paper is as follows. In section II we introduce the
formalism for the calculation of direct detection rates and the annual
modulation signal, and summarize the procedure used by the DAMA/NaI
Collaboration to determine the annual modulation region in the plane
$m_{W}$--$\xi \sigma_{\rm scalar}^{\rm (nucleon)}$ for a purely
spin--independent interacting WIMP. In Section III we describe the
halo models that we intend to discuss and introduce a naming scheme
that will be used throughout the paper.  Section IV is devoted to the
discussion of the constraints on the dark halo of our Galaxy coming
from available observational data.  In Section V the annual modulation
region is calculated in a systematic way for all the models previously
introduced, and the results are discussed.  Finally, Section VI is
devoted to our conclusions.

\section{Direct detection rates and annual modulation effect}
\label{sec:rates}

The expected differential event rate of a WIMP direct search
experiment is given, for a monoatomic detector, by the expression:

\begin{equation}
\frac {dR_{det}}{dE_R}=N_{T}\frac{\rho_0}{m_{W}}
                    \int \,d \vec{w}\,f(\vec w)\,w
                    \frac{d\sigma}{dE_{R}}(w,E_{R}) \label{rate}
\label{eq:diffrate0}
\end{equation}
where $N_T$ is the number of the target nuclei per unit of mass, $m_W$
is the WIMP mass, $\vec w$ and $f(\vec w)$ denote the WIMP velocity
and DF in the Earth frame ($w = |\vec w|$), $d\sigma/dE_R$ is the
WIMP--nucleus differential cross section and $E_R$ is the nuclear
recoil energy. Notice that the detection rate is directly proportional
to the local DM density $\rho_0$. The generalization of
Eq.(\ref{eq:diffrate0}) to a diatomic detector, like NaI, is
straightforward.

The differential cross section is, in general, a sum of a coherent and
a spin--dependent contributions. In this paper we consider only the
case of a WIMP whose interactions are dominated by the coherent
part. In this case the rate may be expressed in terms of the
WIMP--nucleon scalar cross section, $\sigma_{\rm scalar}^{\rm
(nucleon)}$, as:
\begin{equation}
\frac {d\sigma}{dE_R} \simeq \left ( \frac{d\sigma}{dE_R} \right )_{coherent} 
              \simeq \frac{F_n^2(E_R)}{E^{max}_R}
              \left(\frac{1+m_W/m_p}{1+m_W/m_N}\right )^2 A^2
              \sigma^{\rm (nucleon)}_{\rm scalar}, 
\label{eq:diffrate_approx}
\end{equation}

\noindent where $m_p$ and $m_N$ are the proton and nucleus mass,
$A$ is the nuclear mass number, $E_R^{max}$ is the maximal recoil
energy and $F_n(E_R)$ is the nuclear form factor for coherent
interactions which may be parametrized with the usual Helm expression \cite{helm}. 

The relation between the velocities $\vec{v}$ and $\vec{w}$ is given by:
\begin{eqnarray}
&\vec v& = \vec v_{\oplus} + \vec w, \label{eq:change}    \\
&\vec v_{\oplus}& =\vec  v_{\odot} +\vec v_{\oplus rot}, \label{eq:vorb}   
\end{eqnarray}
\noindent where $\vec v_{\oplus}$ and $\vec v_{\odot}$ denote the velocities of the
Earth and the Sun in the Galactic rest frame ($|\vec v_{\odot}|\simeq
v_0 +$12 km/sec), and $\vec v_{\oplus rot}$ is the Earth's orbital velocity
around the Sun ($|\vec v_{\oplus rot}|=$30 km/sec).  Projecting
Eq.(\ref{eq:vorb}) in the galactic plane, one gets:
\begin{equation}
|\vec v_{\oplus}|=|\vec v_{\odot}|+|\vec v_{\oplus rot}|\cos \gamma\cos[\omega
(t-t_0)]
\label{eq:projection}
\end{equation}
\noindent where $\gamma$ is the inclination of the plane of rotation with
respect to the galactic one, $\omega=2\pi/T$ with $T$=1 year, and
$t_0\simeq 2^{nd}$ June corresponds to the day when the Earth's
velocity is at its maximum. 

The change of reference frame of Eqs.
(\ref{eq:change},\ref{eq:vorb},\ref{eq:projection}) introduces through
the DF $f(\vec w)$ a time dependence in the expected rate $R_{det}$.
In order to exploit this time--dependence to extract the modulated
signal from the measured counting rates, we follow the
maximal--likelihood procedure of Ref. \cite{damalast}, to which we
refer for a detailed discussion. This procedure allows to determine
the region in the plane ($m_W$ -- $\xi \sigma_{\rm scalar}^{\rm
(nucleon)}$) which is compatible with the modulation signal. A lower
bound on $m_W$ at the value $m_W = 30$ GeV is applied, to conform to
the analysis of Ref.\cite{damalast,noi_belli}.

The data we analyze in the present paper refer to the full set of data
released so far by the DAMA/NaI Collaboration \cite{damalast}
(DAMA/NaI 0--4), including also the upper limit on $\xi \sigma_{\rm
scalar}^{\rm (nucleon)}$ obtained by the same Collaboration, as
discussed in Ref. \cite{damalast}. The same values of the quenching
factors and of the cut efficiences as in
Refs. \cite{damalast,noi_belli} are used. We stress that also the
determination of upper limits is affected by the choice of the WIMPs
DF.  This means that also when confronting upper limits one has to
specify the galactic halo models which has been considered in the
calculation. The formalism introduced in the present paper may in fact
be used also to quantify the uncertainty in the determination of upper
limits from direct detection experiments.

\section{Halo models}
\label{sec:models}

Let us turn now to the discussion of the galactic halo models and of the
techniques used to calculate the velocity distribution function $f(\vec v)$.
The different models are classified according to the symmetry properties of
their matter density profile (or gravitational potential) and of the velocity
distribution function. We define four classes:

\begin{itemize}
\item[(A)] Spherically symmetric matter density with isotropic velocity
  dispersion;
\item[(B)] Spherically symmetric matter density with non--isotropic velocity dispersion;
\item[(C)] Axisymmetric models;
\item[(D)] Triaxial models.
\end{itemize}
All the models which we describe in this Section, and that will be used in the
rest of the paper, are summarized in Table \ref{tab:models}, where we also
introduce a naming scheme that will be of practical use in the discussion.

\subsection{Spherically symmetric matter density with isotropic velocity dispersion}
\label{sec:classA}
The first class of models is represented by those with a spherically symmetric
matter density $\rho(\vec r)= \rho(r)$ and isotropic velocity distribution
$f(\vec v)= f(v)$. These two conditions imply that the phase--space DF depends
on the space and velocity variables only through the energy, which is an
integral of motion: $F(\vec r,\vec v) = F(\epsilon)$, where
$\epsilon=\Psi(r)-v^2/2$ is the relative energy (per unit mass) of the WIMP and
$\Psi$ is the relative potential, related to the total density $\rho_{tot}$
through the Poisson's equation \cite{binney}:
\begin{equation}
\nabla^2 \Psi=-4 \pi G \rho_{tot}.
\label{eq:poisson}
\end{equation}
Notice that $\rho_{tot}$ refers to all the matter components of the Galaxy,
like the disk, the bulge or the halo, as written in Eq.(\ref{eq:rho_total}).

Once the total potential $\Psi(r)$ is known, the WIMP DF $F(\epsilon)$ may be
worked out by inverting Eq.(\ref{eq:rho_dm}). A change of variables from $r$ to
$\Psi$ allows to cast Eq.(\ref{eq:rho_dm}) as \cite{binney}:
\begin{equation}
4 \pi \int_0^{\Psi} F(\epsilon) \sqrt{2(\Psi-\epsilon)}\;d
\epsilon=\rho_{DM}(\Psi) \, ,
\label{eq:density_psi}
\end{equation}
By performing a Laplace inversion on Eq.(\ref{eq:density_psi}), one
obtains the Eddington formula \cite{binney}:
\begin{equation}
F(\epsilon)=\frac{1}{\sqrt{8} \pi^2}
\frac{d}{d\epsilon}\int_0^{\epsilon}\frac{d\rho_{DM}(\Psi)}{d \Psi}
\frac{d \Psi}{\sqrt{\epsilon-\Psi}}\, .
\label{eq:eddington}
\end{equation}
In Eqs.(\ref{eq:density_psi},\ref{eq:eddington}) the normalization of $\Psi$,
which is defined through Poisson's equation up to an arbitrary constant, is
fixed by requiring that $\Psi(\infty)=0$. The velocity distribution function
$f(v)$ which enters the calculation of direct detection rates is then obtained
as in Eq.(\ref{eq:df}).

Eq.(\ref{eq:eddington}) shows that the dark matter DF $F(\vec r, \vec v)$
depends not only on the halo DM density distribution $\rho_{DM}$, but also on
the density distribution of all the other galactic components. This has to be
the case, since the DM particles move inside the gravitational potential
generated by all the matter that makes up the Galaxy. A complete and rigorous
determination of the DM DF will therefore require to model not only the
galactic halo, but also all the other (disk, bulge) components of the Galaxy
(see, for instance Ref. \cite{dehnen}).

However, WIMP direct detection is directly sensitive only to local properties
of the Galaxy, and in particular the relevant parameters in the calculation of
the detection rate are the local rotational velocity $v_0$ and the local DM
matter density $\rho_0$. In fact, $v_0$ is directly related to the average WIMP
kinetic energy, which is relevant in the scattering process with the nuclei of
the detector, and the change of reference frame of Eqs.
(\ref{eq:change},\ref{eq:vorb}), which is crucial in determining the amount of
annual modulation in the detection rate. Instead, $\rho_0$ is a sort of
normalization factor for the direct detection rate $R_{det}$. The dependence of
$R_{det}$ essentially on local parameters implies that a detailed modeling of
the inner ($r \ll R_0$) part of the Galaxy, where the disk and bulge components
are more relevant and in general dominant over the halo, is not crucial for our
analysis.  Moreover, the matter density of non--halo components at the local
position in the Galaxy ($r=R_0$) are no longer dominant with respect to the
halo matter density at the same position (see, for instance Ref.
\cite{dehnen}). The bulge, in fact, can be described by using a spheroidal
density distribution which gives a sizeable contribution inside the first kpc
from the galactic center, and it is truncated at about $r\simeq $ 2 kpc.  The
disk has an exponential distribution which in most of the models dies away at
about 4 kpc from the galactic center. We will therefore assume in the following
that in the outer Galaxy the dominant contribution to the matter density is
given by the halo:
\begin{equation}
 \rho_{vis}\ll\rho_{DM}\;\;\; {\rm for} \;\;\; r \gsim R_0 \, .
\label{eq:condition}
\end{equation}
The only basic information which is required from the non--halo components is
their contribution to the local rotational velocity $v_0$:
\begin{equation}
v_0^2 = v_{rot}^2(R_0) = \frac{G}{R_0} [M_{vis} + M_{halo}]  \, ,
\end{equation}
where
\begin{equation}
M_{vis~(halo)}\equiv
\int_{r^{\prime}<R_0} d^3 r^{\prime} ~\rho_{vis~(halo)}(r^{\prime}) \, .
\end{equation}
A maximal halo occurs when $M_{vis} \ll M_{halo}$: in this case almost all the
local rotational velocity is supported by the halo and the local DM density
$\rho_0$ gets its maximal value $\rho_0^{max}$ compatible with the given
$v_0$.  The opposite situation occurs when $M_{vis}$ assumes its maximal value
compatible with observations: in this case, the local rotational velocity gets
the maximal contribution from the non--halo components and only a fraction of
$v_0$ is supported by the halo. At the same time $\rho_0$ gets its minimal
value $\rho_0^{min}$, for the same $v_0$.  The constraints on these
parameters are discussed in Section \ref{sec:constraining}.

From the point of view of calculating the DM DF (Eq.(\ref{eq:eddington})), the
occurrence of a maximal or non--maximal halo modifies the gravitational
potential $\Psi(r)$ and therefore the velocity distribution function $f(v)$ is
affected. Indicating with $\Psi_0(r)$ the potential for the maximal halo, the
condition of Eq.(\ref{eq:condition}) allows the generalization to the
non--maximal case as:
\begin{equation}
\Psi(r)=\frac{\rho_0}{\rho_0^{max}}\Psi_0(r)+
\left (1-\frac{\rho_0}{\rho_0^{max}} \right )
\frac{R_0}{r}  ~ v_0^2 \, .
\label{eq:potential_non_maximal}
\end{equation}
The condition of Eq. (\ref{eq:condition}) allows to work out the total
potential for the case of non-maximal halos without explicitly modeling the
visible parts of the Galaxy (bulk, disk): all the dependence of these
components are contained in $M_{vis}$.

As a comment, we notice that the presence of a non negligible contribution to
$v_0$ from the non--halo components alters also the velocity distribution
function of the isothermal sphere. The standard Maxwellian form for the
isothermal sphere is in fact correct only for a maximal halo.

Now that we have discussed the procedure to calculate the velocity distribution
function once the matter density of the DM is given, we proceed to introduce
the different models.

The first type of model is a direct generalization of the isothermal
sphere by introducing a core radius $R_c$ (model A1). The density
profile is:
\begin{equation}
\rho_{DM}(r)=\frac{v_0^2}{4 \pi G} ~ \frac{3 R_c^2+r^2}
{(R_c^2+r^2)^2} \, ,
\label{eq:rho_axi_log_spherical}
\end{equation}
which corresponds to the following potential for a maximal halo:
\begin{equation}
\Psi_0(r)=-\frac{v_0^2}{2}\ln (R_c^2+r^2)\, .
\label{eq:potential_axi_log_spherical}
\end{equation} 
From the analytic form of this potential we will refer to this type of
models as {\em logarithmic models}. The usual isothermal sphere (model
A0) corresponds to the limit $R_c \rightarrow 0$:
\begin{equation}
\rho_{DM}(r)=\frac{v_0^2}{4 \pi G} ~ \frac{1}{r^2} \, ,
\label{eq:rho_isothermal}
\end{equation}
which corresponds to the following potential for a maximal halo:
\begin{equation}
\Psi_0(r)=-\frac{v_0^2}{2}\ln (r^2) \, .
\label{eq:potential_isothermal}
\end{equation} 
For a maximal halo, the velocity distribution function arising from
Eqs.  (\ref{eq:rho_isothermal},\ref{eq:potential_isothermal}) is the
standard Maxwellian one. For a non--maximal halo, a deviation is
present also in the $R_c \rightarrow 0$, as discussed above. The
rotational curve supported by $\Psi_0$ of
Eq. (\ref{eq:potential_axi_log_spherical}) is:
\begin{equation}
v^2_{rot}(r) = v_0^2 ~ \frac{r^2}{(R_c^2 + r^2)} \, .
\label{eq:log_vrot}
\end{equation} 
It is rising for small radii and at large radii becomes flat. In the $R_C
\rightarrow 0$ limit, $v_{rot} = v_0$ for all radii.

A second type of spherical models (A2 and A3) is defined by the
following matter density \cite{evans_pl}:
\begin{equation}
\rho_{DM}(r)=\frac{\beta \Psi_a R_c^{\beta}}{4 \pi G}
~\frac{3 R_c^2+ r^2(1-\beta)}
{(R_c^2+r^2)^{(\beta+4)/2}} \, .
\label{eq:rho_pl_spherical}
\end{equation}
which corresponds to the following potential for a maximal halo:
\begin{equation}
\Psi_0(r)=\frac{\Psi_a R_c^{\beta}}{(R_c^2+r^2)^{\beta/2}}
\;\;\; (\beta\ne 0) \, .
\label{eq:potential_pl_spherical}
\end{equation}
From the analytic form of this potential we will refer to this type of models
as {\em power--law models}. They represent the spherical limit of the more
general class of axisymmetric ``power--law'' models of Ref. \cite{evans_pl}
which will be discussed in Section \ref{sec:classB}. The family of power--law
models given by Eq.(\ref{eq:potential_pl_spherical}) is not defined for
$\beta=0$. However, when $\beta=0$ is substituted in Eq.
(\ref{eq:rho_pl_spherical}), the density of Eq.
(\ref{eq:rho_axi_log_spherical}) is recovered. In fact the logarithmic model
turns out to have the properties of the ``missing'' $\beta=0$ power law
potential. Evaluating Eq.(\ref{eq:rho_pl_spherical}) for $r=R_0$ the parameter
$\Psi_a$ can be expressed in terms of the density $\rho_0$. The rotational
velocity for the power--law models is given by:
\begin{equation}
v_{rot}^2=\frac{\beta \Psi_a
  R_c^{\beta}r^2}{(R_c^2+r^2)^{(\beta+2)/2}} \, ,
\label{eq:vrot_pl}
\end{equation}
and it is asymptotically falling with $r$ if $\beta>0$ (model A2) and
rising if $\beta<0$ (model A3).

The last family of spherical models we consider (models A4--A7) is defined by
the following matter density:
\begin{equation}
\rho_{DM}=\rho_0 \left (  \frac{R_0}{r}\right )^{\gamma}
\left [ 
\frac{1+(R_0/a)^{\alpha}}{1+(r/a)^{\alpha}}
\right ]^{(\beta-\gamma)/\alpha}.
\label{eq:univ_rho}
\end{equation}
for the choice of values of the parameters $\alpha$, $\beta$, $\gamma$
and $a$ summarized in Table \ref{tab:univ_rho}. Except for the Jaffe
model (A4), the other three density profiles (A5, A6, A7) are obtained
from numerical simulations of galaxy evolution.

\subsection{Spherically symmetric matter density with non--isotropic velocity dispersion}
\label{sec:classB}
The procedure described in the previous Section can be generalized to the case
of a non--isotropic velocity distribution, while keeping a spherically
symmetric density profile. In this case, the most general DF is a function of
$\epsilon$ and of the magnitude of the angular momentum vector ($L=|\vec L|$):
$F=F(\epsilon, L)$. Among the different choices which can be performed, a
particularly simple case is obtained when $F$ depends on $\epsilon$ and $L$
only through the so called Osipkov--Merrit variable
\cite{osipkov_merrit,binney}:
\begin{equation}
Q=\epsilon-\frac{L^2}{2 r_a^2} \, ,
\label{eq:om}
\end{equation} 
where the parameter $r_a$ is related to the degree of anisotropy $\beta_0$ of
the velocity dispersion tensor (evaluated at the Earth's position) in the
following way \cite{osipkov_merrit}:
\begin{equation}
\beta_0 = 1-\frac{\bar{v}_{\phi}^2}{\bar{v}_r^2}=\frac{R_0^2}{R_0^2+r_a^2}.
\label{eq:beta0}
\end{equation}
Here the velocity is expressed in spherical coordinates and $\bar{v}_{\phi} =
\bar{v}_{\theta} \ne \bar{v}_r$ (with $\bar{v}^2_i \equiv <v^2_i>-<v_i>^2$,
$i=r, \theta, \phi$).
 
The corresponding DF can be obtained by solving a modified version of
Eddington's inversion formula, which is obtained by making the following
substitutions in Eq.(\ref{eq:eddington}) \cite{binney}:
\begin{eqnarray}
 \epsilon  & \rightarrow & Q \, ,\\
 \rho_{DM}(r) & \rightarrow & \rho_{Q,DM}(r) \equiv  
\left ( 1+\frac{r^2}{r_a^2}   \right )  \rho_{DM}(r) \, .
\end{eqnarray}

The models we consider are the same as discussed in Section
\ref{sec:classA}: the logarithmic model of
Eq. (\ref{eq:rho_axi_log_spherical}) (model B1), the power--law models
of Eq. (\ref{eq:rho_pl_spherical}) (models B2 and B3) and the models
defined by Eq.(\ref{eq:univ_rho}) (models B4--B7). The velocity
distribution functions, which are obtained by solving the Eddington
equation with the Osipkov--Merrit term, are therefore anisotropic with
a degree of anisotropy controlled by the parameter $\beta_0$ related
to $r_a$ as in Eq.(\ref{eq:beta0}).

\subsection{Axisymmetric models}
\label{sec:classC}
In the case of axial symmetry the DF depends in general (at least) on two
integrals of motion, the relative binding energy $\epsilon$ and the component
$L_z$ of the angular momentum along the axis of symmetry. The DF may be written
quite generally as the sum of an even ($F_+$) and an odd ($F_-$) contribution
with respect to $L_z$:
\begin{equation} 
F(\epsilon,L_z) = F_{+}(\epsilon,L_z)+F_{-}(\epsilon,L_z) \, ,
\end{equation}
where:
\begin{equation} 
F_{\pm} = \frac{1}{2}\left [ F(\epsilon,L_z) \pm F(\epsilon,-L_z) \right ] \, .
\end{equation}
When Eq.(\ref{eq:density_psi}) is extended to the axisymmetric case, the
density $\rho_{DM}$ turns out to depend only on the even part $F_{+}$
\cite{binney}, so that, by inverting it, the DF may be determined up to an
arbitrary odd part $F_{-}$. The problem of the determination of $F_{+}$ for an
axisymmetric matter density is both analitycally and numerically hard to
perform and actually it requires a double Laplace inversion on
Eq.(\ref{eq:density_psi}). However, for particular families of axisymmetric
potentials this problem has been solved analytically by Evans
\cite{evans_pl,evans_log}. These families are the axisymmetric generalization
of the first two classes already introduced in Section \ref{sec:classA}: the
first family has a logarithmic potential, the second one has a power--law
potential. We stress that these analytic solution for $F_+$ are obtained under
the assumption that the halo potential is dominant over the other components.
Therefore these solutions correspond to a maximal halo. Even though they do not
represent the most general situation, their simplicity makes them of practical
use and convenient for studying the axisymmetric case.

The first family of axisymmetric potential we consider is the {\em logarithmic
  potential} \cite{evans_log} (models C1 and C2):
\begin{equation}
\Psi_0(R,z)=-\frac{v_0^2}{2}\ln (R_c^2+R^2+\frac{z^2}{q^2}),
\label{eq:potential_axi_log}
\end{equation} 
where $R^2 = x^2 + y^2$ is the radial coordinate along the galactic plane,
$R_c$ is the core radius and $q$ the flatness parameter. The corresponding DM
density is:
\begin{equation}
\rho_{DM}(R,z)=\frac{v_0^2}{4 \pi G q^2}
\frac{(2 q^2+1)R_c^2+R^2+(2-q^{-2})z^2}
{(R_c^2+R^2+z^2q^{-2})^2} \, .
\label{eq:rho_axi_log}
\end{equation}
Eqs.(\ref{eq:potential_axi_log}) and (\ref{eq:rho_axi_log}) are the
axisymmetric generalization of Eqs. (\ref{eq:potential_axi_log_spherical}) and
(\ref{eq:rho_axi_log_spherical}). The corresponding rotational curve is
obtained from Eq.(\ref{eq:log_vrot}) by substituting the radial coordinate $r$
with the radial coordinate in the galactic plane $R$.

By expressing $z$ as a function of $\Psi$ through
Eq.(\ref{eq:potential_axi_log}), the density of Eq.(\ref{eq:rho_axi_log}) may
be decomposed as \cite{evans_pl}:
\begin{equation}
\rho=\rho_0(\Psi)+R^2 \rho_1(\Psi) \, ,
\label{eq:rho_form}
\end{equation}
which allows to determine $F_{+}$ in the form:
\begin{equation}
F_{+}(\epsilon,L_z)=F^0_{+}(\epsilon)+L^2_z F^1_{+}(\epsilon) \, ,
\label{eq:df_analytic}
\end{equation}   
leading to a particularly simple analytic solution for the DF \cite{evans_log}.
We give it for completeness in Appendix \ref{app:evans}.  The relation of
Eq.(\ref{eq:rho_form}) is no longer valid for a non--maximal halo, since in
that case, the change of variable from $z$ to $\Psi$ is not determined by the
potential of Eq.(\ref{eq:potential_axi_log}) but by an axisymmetric analogous
of Eq.(\ref{eq:potential_non_maximal}). Therefore the analytic expression given
by Eq.(\ref{eq:df_analytic}) can be used only for maximal halos, {\em i.e.}
$\rho_0$=$\rho_0^{max}$.

The second family of distribution functions is a generalization of the
axisymmetric logarithmic potential to the case of an asymptotically non--flat
rotational curve, while preserving the property of Eq. (\ref{eq:rho_form}). This is
obtained for the axisymmetric {\em power--law potential}
\cite{evans_pl} (models C3 and C4):
\begin{equation}
\Psi_0(R,z)=\frac{\Psi_a R_c^{\beta}}{(R_c^2+R^2+z^2q^{-2})^{\beta/2}}
\;\;\; (\beta\ne 0).
\label{eq:potential_pl}
\end{equation}
The corresponding matter density is:
\begin{equation}
\rho_{DM}(R,z) = \frac{\beta \Psi_a R_c^{\beta}}{4 \pi G q^2}
~\frac{(2 q^2+1)R_c^2+(1-\beta q^2) R^2+[2-q^{-2}(1+\beta)]z^2}
{(R_c^2+R^2+z^2q^{-2})^{(\beta+4)/2}}.
\label{eq:rho_pl}
\end{equation}
Evaluating Eq.(\ref{eq:rho_pl}) for $R=R_0$, $z=0$ the parameter
$\Psi_a$ can be expressed in terms of the density $\rho_0$.
Eqs.(\ref{eq:potential_pl}) and (\ref{eq:rho_pl}) are the axisymmetric
generalization of Eqs.  (\ref{eq:potential_pl_spherical}) and
(\ref{eq:rho_pl_spherical}) and they posses the same properties
already discussed in Section \ref{sec:classA}.  As for the case of the
logarithmic potential, the rotational curve is obtained from
Eq.(\ref{eq:vrot_pl}) by substituting the radial coordinate $r$ in
Eq.(\ref{eq:vrot_pl}) with the radial coordinate in the galactic plane
$R$, and it is asymptotically falling with $R$ if $\beta>$0 (model C3)
and rising if $\beta<$0 (model C4).

In analogy with the logarithmic case, also for the power--law model an analytic
solution for the DF can be worked out \cite{evans_pl} with the form of Eq. 
(\ref{eq:df_analytic}). This applies again only for a maximal halo. The
analytic formulas for $F_+$ can be found for completeness in Appendix
\ref{app:evans}.

\subsubsection{Co--rotation and counter--rotation of the halo}

As mentioned before, the DF for an axisymmetric model is known up to an
arbitrary odd component $F_{-}$. The DF we summarized above for the Evan's
models all refer to the pure even component: they all have
$F_{-}(\epsilon,L_Z)=0$ and posses no bulk rotation. The case
$F_{-}(\epsilon,L_Z)\ne0$ corresponds to the case of a rotating halo, where the
number of particles moving clockwise around the axis of symmetry is different
from that in the opposite sense.

A family of DF's with bulk rotation can be studied by constructing an explicit
example for $F_{-}$. This can be done, starting from a generic $F_{+}$, by
considering the linear combination \cite{lynden_corot,kamion_corot,noi_corot}:
\begin{equation}
F_{-}(\epsilon,L_z)=F_{right}(\epsilon,L_z)-F_{left}(\epsilon,L_z) \, ,
\end{equation}
where

\begin{equation}
F_{right}(\epsilon,L_z)=\left\{ \begin{array}{ll}
            F_{+}(\epsilon,L_z), & v_{\phi}>0 \, , \\
            0,                 & v_{\phi}<0 \, , 
           \end{array} \right.
\end{equation}
and
\begin{equation}
F_{left}(\epsilon,L_z)=\left\{ \begin{array}{ll}
            0,                   & v_{\phi}>0, \\
            F_{+}(\epsilon,L_z),   & v_{\phi}<0, 
           \end{array} \right. 
\end{equation} 
The distributions $F_{right}$ and $F_{left}$ describe the configurations with
maximal $|\bar{v}_{\phi}|$ with the same density profile as $F_{+}$
\cite{lynden_corot}.  A DF with an intermediate value of $\bar{v}_{\phi}$ can
be obtained as a linear combination of $F_{+}$ and $F_{-}$, or, equivalently,
of $F_{left}$ and $F_{right}$:
\begin{equation}
F(\epsilon,L_z)= \eta F_{right}(\epsilon,L_z)+(1-\eta) F_{left}(\epsilon,L_z).
\label{eq:corot}
\end{equation}
The parameter $\eta$ ranges from 1 (maximal co--rotation) to 0 (maximal
counter--rotation) and is related to the dimensionless spin parameter $\lambda$
of the Galaxy by: $\lambda=0.36 |\eta-0.5|$ \cite{kamion_corot}. In order to be
consistent with the available extensive numerical work on galaxy formation,
$\lambda$ should not exceed the value 0.05 \cite{spingalaxy}, implying
$0.36\lsim \eta \lsim 0.64$. For all the Evans model discussed in this Section,
we will also study the co-- and counter-- rotating situations, adopting the two values
$\eta=0.36$ and $\eta=0.64$

\subsection{Triaxial models}
\label{sec:classD}
The last class of models we wish to discuss is represented by the {\em triaxial
  potential} discussed in Ref. {\cite{triaxial}}:
\begin{equation}
\Psi_0(x,y,z)=-\frac{1}{2} v_0^2 \ln \left (x^2+\frac{y^2}{p^2}+
\frac{z^2}{q^2}\right),
\label{eq:potential_triaxial}
\end{equation}
which, for a maximal halo, corresponds to the DM density:
\begin{equation}
\rho_{DM}(x,y,z)=\frac{v_0^2}{4 \pi G}
\frac{A x^2+B y^2/p^2+C z^2/q^2}
{\left (x^2+ y^2/p^2+z^2/q^2\right )^2}
\label{eq:rho_triaxial}
\end{equation}
where $A=p^{-2}+q^{-2}-1$, $B=1+q^{-2}-p^{-2}$ and $C=1+p^{-2}-q^{-2}$. In Ref.
\cite{triaxial} the velocity DF $f(\vec v)$ of the system is approximated by a
triaxial Gaussian with semiaxes equal to the velocity dispersions as obtained
by the solutions of the Jeans equations:
\begin{eqnarray}
\bar{v}^2_r & = & \frac{v_0^2}{(2+\delta)(p^{-2}+q^{-2}-1)} \label{eq:sigma_major}\\
\bar{v}^2_{\phi} & =& \frac{v_0^2 (2 q^{-2}-1)}{2 (p^{-2}+q^{-2}-1)} \\
\bar{v}^2_{\theta} & = & \frac{v_0^2 (2 p^{-2}-1)}{2 (p^{-2}+q^{-2}-1)}
\end{eqnarray}
when the Earth's position is on the major axis of the
equipotential ellipsoid (models D1 and D2), and:
\begin{eqnarray}
\bar{v}^2_r & = & \frac{v_0^2 p^{-4}}{(2+\delta)(1+q^{-2}-p^{-2})} \label{eq:sigma_inter}\\
\bar{v}^2_{\phi} & = & \frac{v_0^2 (2 q^{-2}-p^{-2})}{2(1+q^{-2}-p^{-2})}\\
\bar{v}^2_{\theta} & = &  \frac{v_0^2 (2-p^{-2})}{2(1+q^{-2}-p^{-2})}
\end{eqnarray} 
when the Earth's position is on the intermediate axis (models D3 and D4).
In Eqs.(\ref{eq:sigma_major},\ref{eq:sigma_inter}) the quantity $\delta$
is a free parameter that in the spherical limit ($p=q=1$)
quantifies the degree of anisotropy of the velocity dispersion tensor:
\begin{equation}
\frac{\bar{v}_{\phi}^2}{\bar{v}_r^2}=\frac{2+\delta}{2}.
\label{eq:delta}
\end{equation}

\section{Constraining the models}
\label{sec:constraining}

Once a given model is chosen for the velocity distribution function of the dark
matter particles, the parameters of the model have to be fixed using
observational data. Unfortunately, due to its ``darkness'', all our knowledge
of the halo is of indirect nature \cite{turner,dehnen} and it includes
requirements on the circular rotational speed (constraints on its flatness and
its value at the solar circle and in the outer regions of the Galaxy) as well
as observational constraints on the local surface density of the disk and on
the dispersion velocity of the bulge. In general, one should construct a
composite model of the Galaxy where the DM is coupled to other components like
the disk and the bulge, and the parameters describing the various components
are varied independently requiring that the observational constraints are
satisfied.

We have already noticed that WIMP direct detection rates are particularly
sensitive to the value of the rotational velocity $v_0$ and the local DM
density $\rho_0$ (both evaluated at the solar circle). The procedure we follow
in order to determine the allowed ranges for $v_0$ and $\rho_0$ in each
galactic model is explained in the following. First of all, the experimental
informations we use are: the allowed range for the local rotational velocity,
the amount of flatness of the rotational curve of our Galaxy and the maximal
amount of non-halo components in the Galaxy. The first information directly
fixes the allowed interval for $v_0$, irrespective of the galactic halo model.
The other constraints are used in order to determine the allowed ranges for
$\rho_0$, for each halo model.

The allowed interval for $v_0$ is:
\begin{equation}
v_0=(220 \pm 50)\;\; {\rm km \;\; sec^{-1}} \;\;\;\;\; {\rm (90 \%
  \;\; C.L.}),
\label{eq:v0}
\end{equation} 
which conservatively relies on purely dynamical observations \cite{kochanek}.
Proper motion measurements of nearby stars \cite{cepheids} lead to similar
estimates for the central value of $v_0$, with a significantly smaller
uncertainty. However they are based on the assumption of circular orbit of the
observed objects.  For definiteness, we will use in the following three
representative values for $v_0$, which correspond to its central value and to
the boundaries of its allowed 90\% C.L. range of Eq.(\ref{eq:v0}): $v_0 = 170,
220, 270$ km sec$^{-1}$.

For the three representative values of $v_0$ we then determine the
corresponding allowed ranges for $\rho_0$.  For each halo model and for each
value of $v_0$, we calculate, as a function of $\rho_0$, two quantities: {\it
  (i)} the total amount of mass $M_{vis}$ in components other than the halo
({\em e.g.  }: disk, bulge) which is necessary in order to match the given
value of local rotational velocity $v_0$; {\it (ii)} the value of the
rotational curve at a distance of 100 kpc from the center of the Galaxy:
$v_{rot}^{100}\equiv v_{rot}$ ($R=100$ kpc). These two quantities are somewhat
constrained from observations, even though their constraints are often obtained
by using some degree of galactic modeling. We conservatively quote the following
ranges \cite{turner,dehnen}:
\begin{eqnarray}
1 \cdot 10^{10} M_{\odot} & \lsim  M_{vis}  \lsim & 6 \cdot 10^{10} M_{\odot} \label{eq:lim_mvis}\\
0.8 \cdot v_0 & \lsim  v_{rot}^{100}  \lsim & 1.2 \cdot v_0. \label{eq:lim_vrot}
\end{eqnarray}
where $M_{\odot}$ denotes the solar mass. The first constraint limits the
amount of non-halo components, while the second is a constraints of
``essentially flatness'' rotational curve: only galactic halo models which
provide a rotational curve which does not deviate from a flat one more than
20\% at 100 kpc are accepted. The constraint of Eq.(\ref{eq:lim_vrot}) is
compatible with the estimates of the galactic mass at large radii as obtained
by the dynamics of satellites of the Galaxy \cite{dehnen}.

The behaviour of $M_{vis}$ and $v_{rot}^{100}$ as a function of $\rho_0$ is
shown if Fig. 1 (which refers to $v_0 = 220$ km s$^{-1}$), Fig. 2 ($v_0 = 170$
km s$^{-1}$) and Fig. 3 ($v_0 = 270$ km s$^{-1}$). Each line refers to a
different halo model. The upper panels show that $M_{vis}$ is a decreasing
function of $\rho_0$, since increasing the amount of dark matter in the Galaxy
implies that less matter in other galactic components is required to support
the rotational curve. On the other hand, the value of the rotational velocity
in the outer Galaxy is totally supported by the dark halo, and it is larger for
more massive haloes.

When the constraints expressed in Eqs. (\ref{eq:lim_mvis}) and
(\ref{eq:lim_vrot}) are simultaneously applied, an allowed interval for
$\rho_0$ may be derived for each halo model.
 
The procedure outlined above may be used as a simple recipe for
identifying the intervals for the local density parameter. However,
some caution must be taken in the application of the bounds on
$M_{vis}$ in Eq.(\ref{eq:lim_mvis}).  As already discussed in the
previous Section, for the models of class A and class B we solve the
Eddington equation in order to determine the velocity distribution
function of dark matter particles. In this case we can take into
account also the situation in which the local rotational velocity is
only partially supported by the halo, and therefore we can apply the
limits to $M_{vis}$ given by Eq.(\ref{eq:lim_mvis}).  Instead, the
analytic models of class C and class D may be applied only to the
extreme case of a fully maximal--halo, since the analytic formulae for
$f(\vec v)$ are derived under the assumption that only the halo matter
density is present. For these models (class C and class D) the only
case we can deal with is that of a fully maximal halo, which
corresponds to $M_{vis}=0$. This represents a conservative upper limit
for $\rho_0$. From Fig.  1--3 we can see that upper bounds to $\rho_0$
obtained from the $M_{vis}=0$ limit are only a few percent larger than
what is obtained by imposing the lower limit of
Eq.(\ref{eq:lim_mvis}): $M_{vis}=1 \cdot 10^{10} M_{\odot}$. For
consistency and simplicity, we will use $M_{vis} \simeq 0$ as a lower
limit also for models of class A and B.

The allowed intervals for $\rho_0$ that we obtain by imposing the bounds on
$M_{vis}$ and $v_{rot}^{100}$ are listed in Table \ref{tab:intervals}.  As
discussed above, both values $\rho_0^{min}$ and $\rho_0^{max}$ will be used in
the next Sections to perform the modulation analysis of the DAMA/NaI
experimental data, while only $\rho_0^{max}$ will be used for the analysis of
the data for models of class C and D. We stress that the reason for this stands
in the fact that for the analytic models of classes C and D, we know $f(\vec
v)$ only for a maximal--halo.

\section{Results and discussion}
\label{sec:discussion}

In this Section we make use of the halo models described in Section III, with
the choice of parameters shown in Table \ref{tab:models}, to analyze the
annual--modulation signal present in the DAMA/NaI data \cite{damalast} in terms
of relic WIMPs with purely coherent interactions. The procedure is the one
outlined in Section II. The results are presented as 3$\sigma$
annual--modulation regions shown in the plane $\xi \sigma_{\rm scalar}^{\rm
  (nucleon)}$ versus $m_{W}$.  All figures are divided in three panels which
correspond to the cases: $v_0=170$, 220, 270 km sec$^{-1}$.

A general feature of all the models is that by raising the parameter $v_0$ the
modulation region moves from the upper--right to the lower--left of the
$m_W$--$\xi \sigma_{\rm scalar}^{\rm (nucleon)}$ plane.  This is easily
understood since, for a given DM density profile, higher values of $v_0$ imply
higher values of $\rho_0$ (through Eq. (\ref{eq:vrot})) and of the velocity
ellipsoid $\sigma_{ij}\equiv <v_i v_j>$ (through the Jeans equations); the
experimental value of the signal and the measured WIMP--nucleus recoil energy
are fixed by the data, therefore the modulation region moves downward because:
\begin{equation}
\frac{dR_{det}}{dE_R} \propto \rho_0 \times \sigma_{\rm scalar}^{\rm (nucleon)}
\end{equation}
(see Eq.  (\ref{eq:diffrate0})) and moves to lower masses because:
\begin{equation}
E_R \propto m_W \times <v^2>.
\end{equation}

On the other hand, when one compares the different density profiles that we
have discussed in the previous Sections, it is worth noticing that, for a given
value of $v_0$, a stronger singularity in the galactic center lowers the value
of $\rho_0^{max}$ (in order to keep constant the mass integral) with the
consequence that the lower part of the modulation region rises. So the smallest
values of $\xi \sigma_{\rm scalar}^{\rm (nucleon)}$ are reached by the models
with a less singular density profile.

\begin{figure} \centering
  \resizebox{\textwidth}{!}{\includegraphics*[18pt,144pt][592pt,640pt]{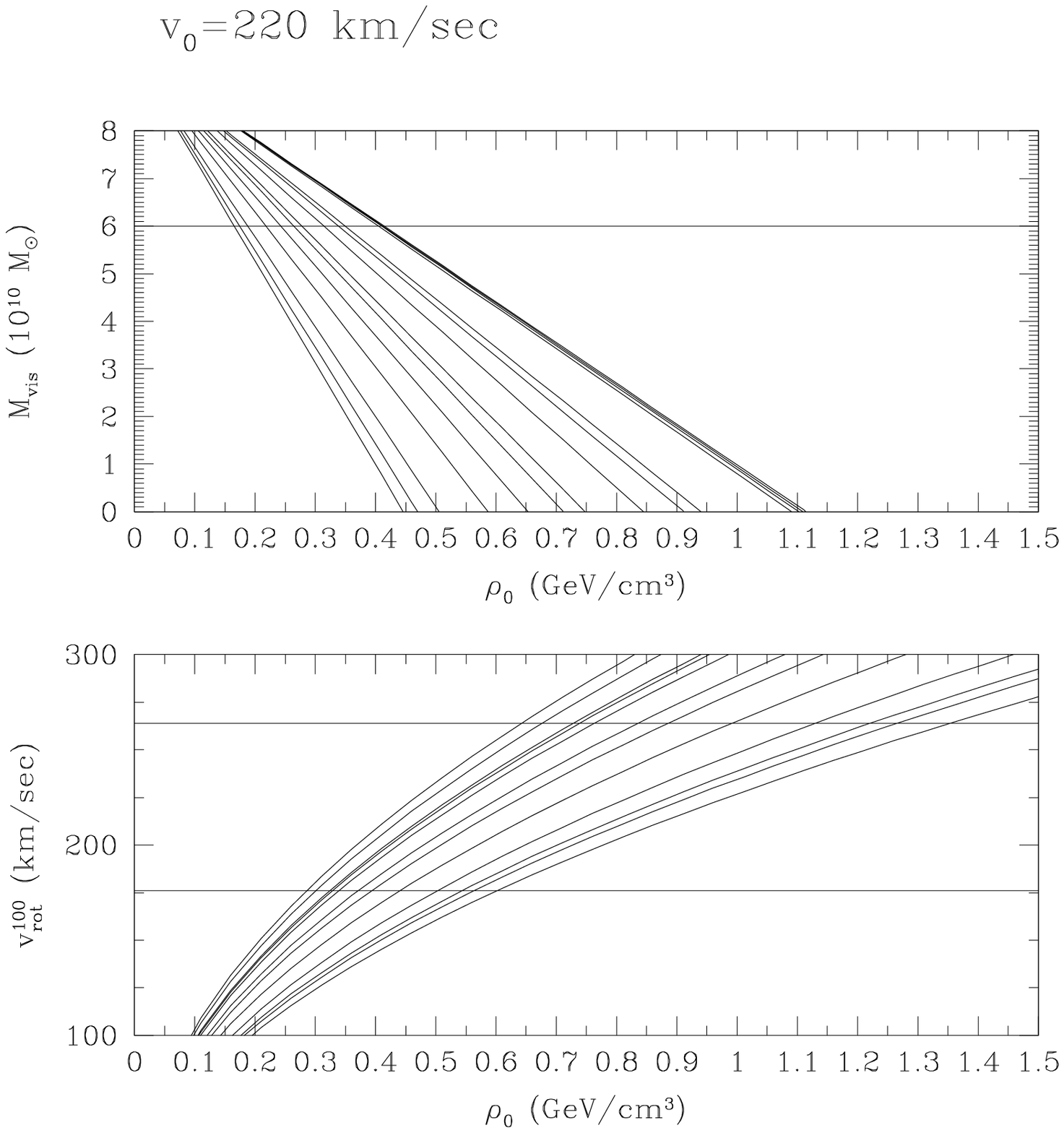}}
  \vspace{-70pt}
  \caption{The quantities $M_{vis}$ (upper panel) and $v_{rot}^{100}\equiv
    v_{rot}(R = 100 ~{\rm kpc})$ (lower panel) calculated as a function of
    $\rho_0$ for the halo models summarized in Table \ref{tab:models} and for
    $v_0$=220 km sec$^{-1}$.  The different curves correspond, from left to
    right, to the following halo models: A4, A0, D3, A3, A6, A1, A5, D1, A7,
    C1, C4, C3, A2, C2 (upper panel); A3, A0, D3, A5, A1, A6, A2, A4, C3, A7,
    C2, C1 (lower panel: here C4 and D1 are not plotted because
    indistinguishable from A7). The horizontal lines indicate the constraints
    discussed in Section \ref{sec:constraining}. The B1-7 models have the
    same density distribution of the corresponding A1-7 models. D2 and D4 have the
    same distribution as D1 and D3, respectively.
\label{fig:check_densities_220}}
\end{figure}

\begin{figure} \centering
  \resizebox{\textwidth}{!}{\includegraphics*[18pt,144pt][592pt,640pt]{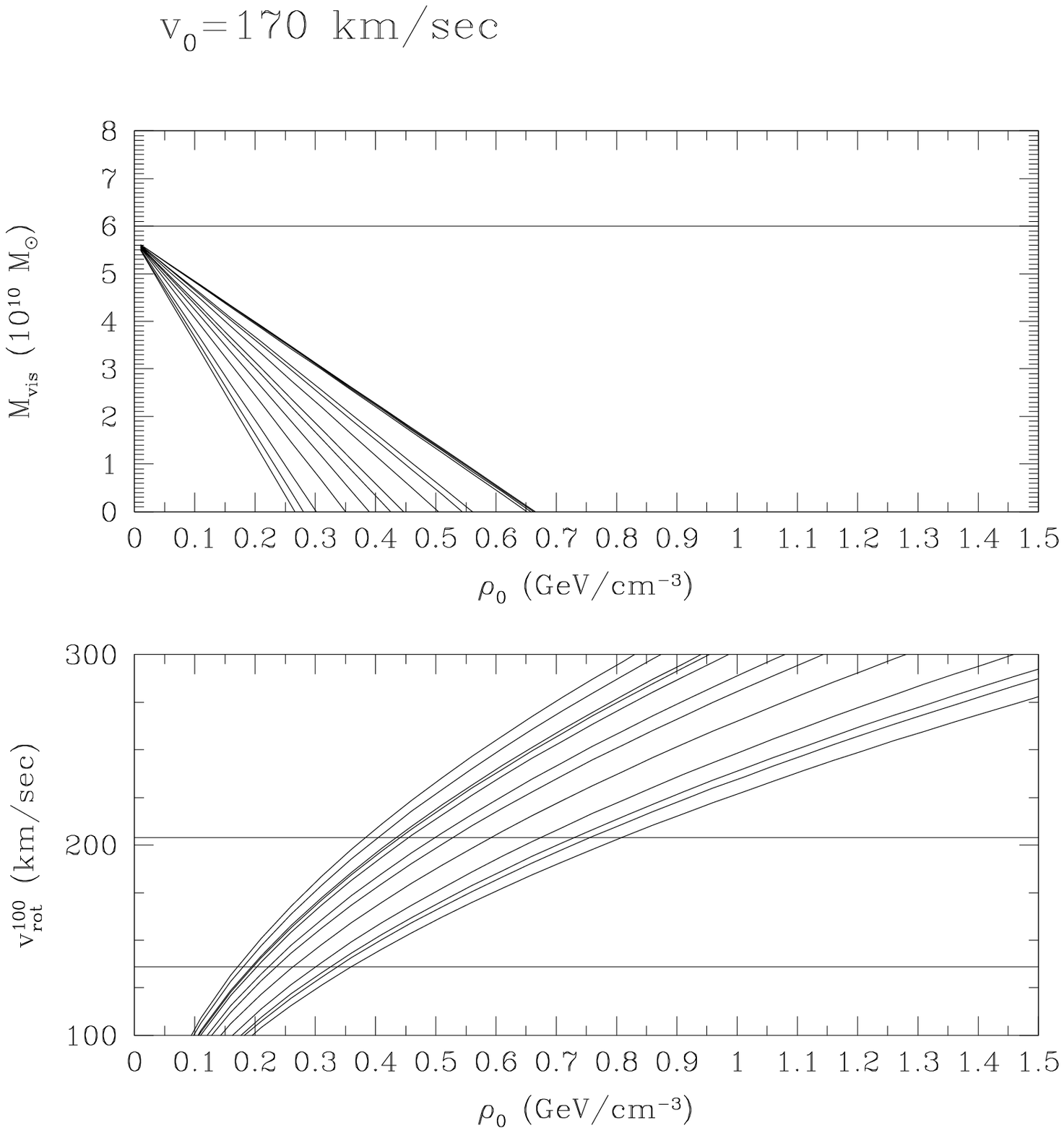}}
  \vspace{-70pt}
  \caption{
    The same as in Figure \ref{fig:check_densities_220} for $v_0=170$ km
    sec$^{-1}$ (the ordering of all the different curves is maintained).
\label{fig:check_densities_170}}
\end{figure}

\begin{figure} \centering
  \resizebox{\textwidth}{!}{\includegraphics*[18pt,144pt][592pt,640pt]{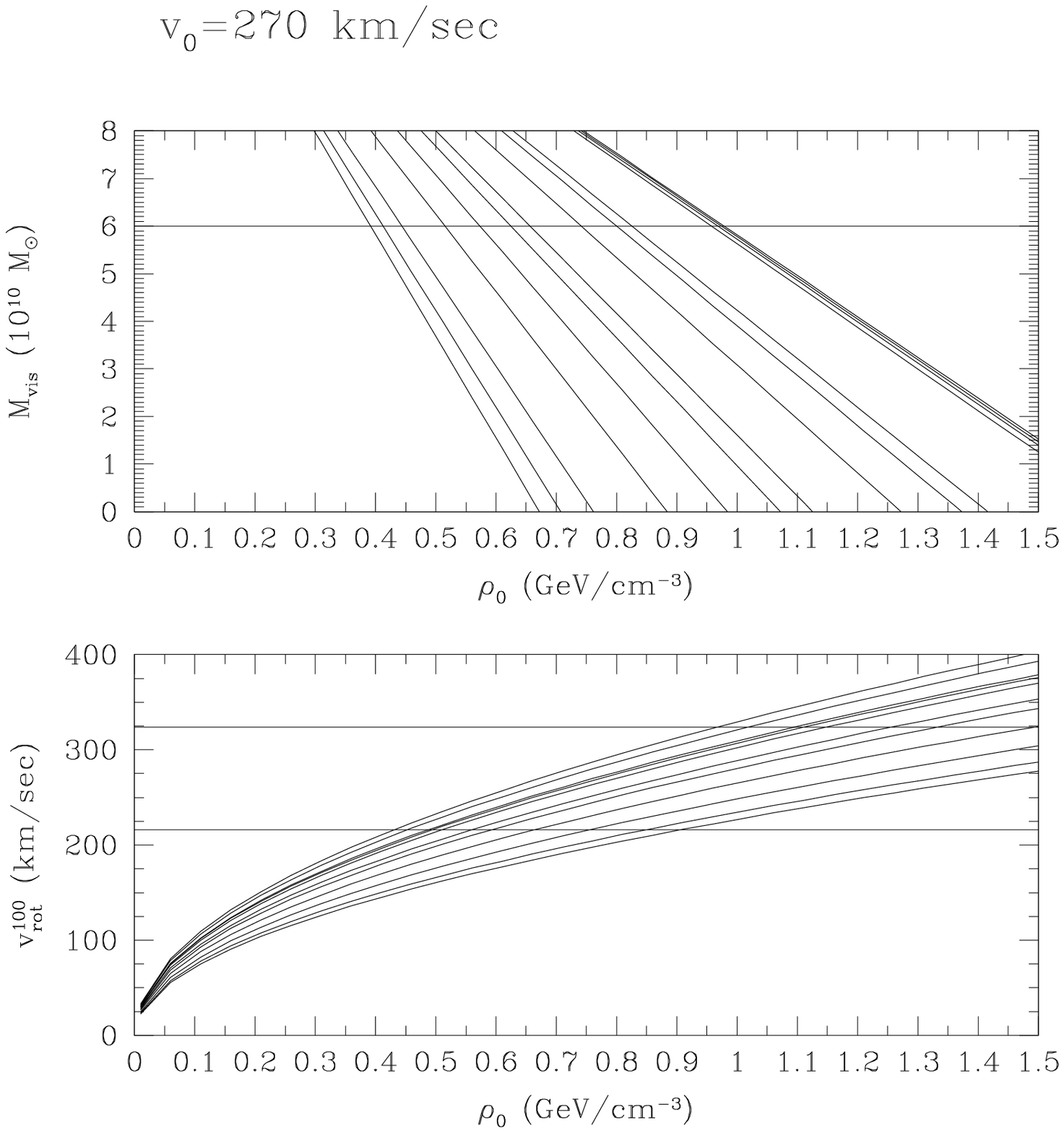}}
  \vspace{-70pt}
  \caption{
    The same as in Figure \ref{fig:check_densities_220} with $v_0=270$ km
    sec$^{-1}$ (the ordering of all the different curves is maintained).
\label{fig:check_densities_270}}
\end{figure}

Figures 4--18 show the result of the analysis for the models with a spherically
symmetric density profile (models A0--7, B1--7).  Models A0--7 have an
isotropic velocity dispersion, while in models B1--7 a degree of anisotropy in
the velocity dispersion is introduced through the Osipkov--Merrit term of
Eq.(\ref{eq:om}). The amount of anisotropy is controlled by the $\beta_0$
parameter of Eq. (\ref{eq:beta0}) and it has been fixed at the value:
$\beta_0=0.4$. This value corresponds to a radial anisotropy. In order to
account for the uncertainty in the DM local density $\rho_0$, for each case two
regions are given, the higher (lower) corresponding to $\rho_0=\rho_0^{min}$
($\rho_0^{max}$), where $\rho_0^{min}$ and $\rho_0^{max}$ have been obtained as
discussed in Section IV and are reported in Table \ref{tab:intervals}.  The
numerical values of the parameters summarized in the third column of Table
\ref{tab:models} have been chosen in order to ensure the compatibility of the
curves of Figs.  \ref{fig:check_densities_220}, \ref{fig:check_densities_170},
\ref{fig:check_densities_270} with the constraints of
Eqs.(\ref{eq:lim_mvis},\ref{eq:lim_vrot}), discussed in Section
\ref{sec:constraining}. The Jaffe models A4, B4, which goes into the isothermal
sphere when $R_c \rightarrow \infty$, have been calculated for the smallest
allowed value of the core radius $R_c$ in order to examine the case of maximal
departure from the usual scenario. As a consequence of this, for this model
$\rho_0^{min} \simeq \rho_0^{max}$ and in Figs.  \ref{fig:a4},\ref{fig:b4} the
upper and lower modulation regions are superimposed.

The effect of radial anisotropy in the velocity dispersion tensor
($\bar{v}_r>\bar{v}_{\theta}=\bar{v}_{\phi}$), which occurs for the models of
class B, may be seen by comparing Figs. \ref{fig:a1}--\ref{fig:a7} with the
corresponding Figs.  \ref{fig:b1}--\ref{fig:b7}. As a general feature, a
reduction of the modulation effect is expected, since the WIMPs phase space is
depopulated along the direction of the Sun's velocity.  This is confirmed by
the fact that in most cases the modulation regions move upwards and widens,
although the size of the effect can be small.  The effect of radial anisotropy
on the WIMP mass is more involved.  In particular, the modulation regions for
models B1, B3, B5, B6 extend to heavier WIMP masses compared to the
corresponding isotropic cases, while for models B2, B4, B7 the region moves to
smaller WIMP masses.

As already pointed out, for the models belonging to classes C and D only the
regions for $\rho_0$=$\rho_0^{max}$ are shown.  As far as the axisymmetric
models of class C are concerned, they are shown in Fig. \ref{fig:c1},
\ref{fig:c2}, \ref{fig:c3}, \ref{fig:c4}. In each model the flatness parameter
$q$ has been chosen in order to have the maximal flatness compatible with
observations and with the positivity of the DF. The main effect induced by
flatness is through the increase in the local density $\rho_0$, as can be seen
in Table \ref{tab:intervals}.  As a consequence, the modulation regions for
these models reach values of $\xi \sigma_{\rm scalar}^{\rm (nucleon)}$
significantly below the levels of the spherical cases. In Figs.
\ref{fig:c1_corot}, \ref{fig:c2_corot}, \ref{fig:c3_corot}, \ref{fig:c4_corot}
the same models are shown with a co--rotation effect of the halo implemented
through Eq.(\ref{eq:corot}) with $\eta=0.64$, while in Figs.
\ref{fig:c1_controrot}, \ref{fig:c2_controrot}, \ref{fig:c3_controrot},
\ref{fig:c4_controrot} for the same models a counter--rotation effect with
$\eta=0.36$ has been introduced.  The main consequence of halo co--rotation is
a decrease of the relative velocity between WIMPs and the Earth. The energy of
WIMPs is therefore smaller and in order to produce the same recoil energy in
the detector the WIMPs have to be heavier. This can be verified in the figures,
where the modulation region of co--rotating models may reach very high WIMP
masses, even higher than 200 GeV.  By the same token, in counter--rotating
models the modulation region is shifted toward lower masses.

The peculiar shape of the modulation region of Fig. \ref{fig:c1_corot} deserves
some comments. The two disconnected closed contours, which arise at different
$m_W$ values, are indicative of the superposition in the WIMP phase space of
two components with well separated r.m.s. velocities.  This is exemplified in
Fig.  \ref{fig:test_c1} for the models B1 ($q=1$) and C1 ($q=1/\sqrt 2$), where
the contour plots of the corresponding DF's are plotted in the $v$--$v_{\phi}$
plane (in the galactic rest frame) and in the $w$--$w_{\phi}$ plane (in the
Earth's rest frame).  It is evident from the figure that, in flattened models,
small $|L_z|$ orbits are depopulated compared to the spherical case, leading to
two well separated populations with $v_{\phi}>0$ and $v_{\phi}<0$.  These two
components have the same temperature in the galactic rest frame, but develop
different r.m.s. velocities when boosted in the Earth's rest frame. As shown in
Fig. \ref{fig:test_c1}, the relative weight of the two populations can be
tilted towards slow or fast WIMPs in the two cases of a co--rotating or
counter--rotating halo, respectively. This explains why the co--rotating model
of Fig. \ref{fig:c1_corot} develops a second minimum at high WIMP masses, and
this is not observed in the corresponding non--rotating or counter--rotating
cases. We have numerically verified that the peculiar disconnected region at
high WIMP masses reduces in size when the flatness parameter is increased
(since in this way the velocity distribution function becomes more similar to
the non--flattened one), while it shifts towards lower masses when the core
radius $R_c$ is decreased.

We conclude the discussion of our results with the triaxial models shown in
Figs. \ref{fig:d1}, \ref{fig:d2}, \ref{fig:d3}, \ref{fig:d4} where, to be
definite, the same choice of parameters of Ref.\cite{triaxial} is adopted. For
these models a general solution for the DF is not available. Only the velocity
ellipsoid of Eq. (\ref{eq:sigma_major}--\ref{eq:sigma_inter}) is known, and it
is used to fix the second moments of a non--isotropic Maxwellian. This explains
why the shape of the modulation regions is quite similar to the standard case.
In models D1 and D2 the Earth is assumed to be located on the major axis of the
density ellipsoid, while in models D3 and D4 it is placed on the intermediate
axis. Since in the two cases the Sun's position $R_0$ is the same, $\rho_0$ is
higher for models D1, D2 than for models D3, D4 (see Table
\ref{tab:intervals}).  As a consequence, the modulation regions of Fig.
\ref{fig:d1}, \ref{fig:d2} reach smaller values of $\xi \sigma_{\rm scalar}^{\rm
  (nucleon)}$ compared to those of Figs.  \ref{fig:d3}, \ref{fig:d4}.  Models D1
and D3 (D2 and D4) have $\delta=-1.78$ ($\delta=16$), which implies a radial
(tangential) anisotropy of the velocity ellipsoid (see Eq. (\ref{eq:delta})).
Solving the Jeans equation for the potential of
Eq.(\ref{eq:potential_triaxial}) in the spherical limit $q=p=1$ (which
corresponds to a non-cored isothermal sphere) leads to the relation
\cite{binney}:
\begin{equation}
v_{\phi}^2+v_{\theta}^2=\frac{3}{2} v_0^2.
\label{eq:jeans}
\end{equation}
Note that this property is a consequence of the flatness of the rotational
curve, and would not be true, for instance, including a core radius in the
potential.  Eq.(\ref{eq:jeans}) implies that, in this model, the tangential
components of the velocity ellipsoid are fixed by $v_0$. As a consequence of
this, radial anisotropy ($v_r>v_{\theta}=v_{\phi}$) corresponds to faster WIMPs
and tangential anisotropy to slower WIMPs. This shifts the modulation regions
towards smaller values of $m_W$ in Figs. \ref{fig:d1}, \ref{fig:d3} and higher
values of $m_W$ in Figs. \ref{fig:d2}, \ref{fig:d4}. This effect is sizeable in
the case of tangential anisotropy, where the modulation region may extend up to
$m_W \simeq 270$ GeV.

The results of this section are summarized in Fig. \ref{fig:all}, where all the
modulation regions previously discussed have been plotted jointly. A
convolution of all the regions may be indicative of the uncertainties in the
determination of the WIMP modulation signal due to the modeling of the WIMP DF.
As a final result, we show such a convolution in Fig. \ref{fig:convolution},
where a single curve in the plane $m_W$--$\xi \sigma_{\rm scalar}^{\rm
  (nucleon)}$ is plotted by collecting the information contained in the
analyses of all the non--rotating models considered in this paper. The region
is compared with the original annual modulation region obtained in Ref.
\cite{damalast} for an isothermal sphere model of the galactic halo with
rotational velocity $v_0 = 220$ km s$^{-1}$ and local dark matter density
$\rho_0 = 0.3$ GeV cm$^{-3}$. From Fig.\ref{fig:convolution} we see that the
DAMA/NaI annual modulation result is compatible with WIMPs masses up to $m_W
\simeq 270$ GeV and WIMP--nucleon cross sections in the interval: $10^{-10}
~{\rm nbarn} \lsim \xi \sigma_{\rm scalar}^{\rm (nucleon)} \lsim 6\times
10^{-8} ~{\rm nbarn}$ when the uncertainties in the WIMP velocity DF is taken
into account. Co--rotating models with maximal corotation can extend the mass
range even further, up to $m_W \simeq$ 500--900 GeV, for cross section of the
order ${\rm few} \times 10^{-9} ~{\rm nbarn} \lsim \xi \sigma_{\rm scalar}^{\rm
  (nucleon)} \lsim 2 \times 10^{-8} ~{\rm nbarn}$, as it can be seen, for
instance, in Fig.  \ref{fig:all}.

\section{Conclusions}
\label{sec:conclusions}

In the present paper we have extended a previuos analyses of the DAMA
modulation experiment for the case of a WIMP with a purely spin-independent
coupling, by discussing in detail the implications on the results of the
uncertainties on the dark matter galactic velocity distribution. We have
studied a large number of viable models which deviate from the standard
isothermal sphere in the matter density profile, in the presence of
anisotropies of the velocity dispersion tensor and in effects of rotation of
the galactic halo. The different models have been classified according to the
symmetry properties of their matter density profile (or gravitational
potential) and of the velocity distribution function. We have specifically
considered: ({\em a}) spherically symmetric matter density with isotropic
velocity dispersion; ({\em b}) spherically symmetric matter density with
non--isotropic velocity dispersion; ({\em c}) axisymmetric models; ({\em d})
triaxial models.

The different models have then been used to re--analyze the DAMA/NaI 0--4 data
collected by the DAMA/NaI Collaboration \cite{damalast}; in particular a total
exposure of 57986 kg day, which corresponds to 4 annual cycles, has led to the
observation of an annual modulation effect. The hypothesis of WIMP annual
modulation, already favoured in the previous studies \cite{damalast,noi_belli}
by using an isothermal sphere, is confirmed in all the investigated scenarios,
and the effects of the different halo models on the determination of the
allowed maximum--likelihood region in the WIMP mass and WIMP--nucleon
cross--section have been derived. We can summarize that the DAMA/NaI annual
modulation result is compatible with WIMPs masses up to $m_W \simeq 270$ GeV
and WIMP--nucleon cross sections in the interval: $10^{-10} ~{\rm nbarn} \lsim
\xi \sigma_{\rm scalar}^{\rm (nucleon)} \lsim 6\times 10^{-8} ~{\rm nbarn}$,
when the uncertainties in the WIMP velocity DF is taken into account.  When
also co--rotation of the galactic halo is considered, the mass range extends
further to $m_W \simeq$ 500--900 GeV, for cross section of the order ${\rm few}
\times 10^{-9} ~{\rm nbarn} \lsim \xi \sigma_{\rm scalar}^{\rm (nucleon)} \lsim
2 \times 10^{-8} ~{\rm nbarn}$.  These intervals quantify the extent of the
annual modulation region for WIMPs with purely spin--independent couplings, as
due to uncertainties in the phase space distribution function of galactic WIMPs.

\section*{Acknowledgements}
We thank Prof. R. Bernabei and Prof. A. Bottino for useful discussions and
suggestions. We are indebted with the DAMA/NaI Collaboration for allowing us to
use its data for this investigation.

\appendix
\section{Axisymmetric models}
\label{app:evans}

In this appendix we give, for completeness, the analytic formulas of the DF's
for the axisymmetric potentials of Eqs.
(\ref{eq:potential_axi_log},\ref{eq:potential_pl}) adapted from Refs.
\cite{evans_log,evans_pl}. All expressions are written in the reference frame
of the Galaxy.

\subsection{Logarithmic potential}

The DF for the logarithmic potential of Eq. (\ref{eq:potential_axi_log}) can be
written as:
\begin{equation}
F(\epsilon,L_z^2)=( A+B) \exp \left ( \frac{-2 v^2}{v_0^2} \right )+
C \exp \left ( \frac{- v^2}{v_0^2} \right ) \, ,
\end{equation}
where:
\begin{eqnarray}
A&=& F_0 \; 4 \pi \left (\frac{2}{\pi}\right )^{5/2}\left (
  \frac{v_{\phi}}{v_0}\right)^2
\frac{R_0^4}{(R_0^2+R_c^2)^2}
\frac{1-q^2}{q^2} \, ,\\
B&=& F_0 \; 4 \pi \left (\frac{2}{\pi^5}\right )^{1/2} \frac{R_0^2
  R_c^2}{(R_0^2+R_c^2)^2} \frac{1}{q^2} \, ,\\
C&=& F_0 \; \pi^{-3/2}\; \frac{R_0^2}{R_0^2+R_c^2} \frac{2 q^2-1}{q^2} \, ,
\end{eqnarray}
and $F_0=0.47$ GeV cm$^{-3}/ v_0^3$.

\subsection{Power--law potential}

The DF for the power--law potential of Eq. (\ref{eq:potential_pl}) for
$\beta>0$ can be written as:
\begin{equation}
F(\epsilon,L_z^2)=A \tilde{\epsilon}^{4/\beta-3/2}+
B \tilde{\epsilon}^{4/\beta-1/2}+
C \tilde{\epsilon}^{2/\beta-1/2} \, ,
\label{eq:dfpl}
\end{equation}
where:
\begin{equation}
\tilde{\epsilon} \equiv \frac{\epsilon}{\Psi_a}=\frac{\Psi-\frac{1}{2}
  v^2}{\Psi_a}=\zeta^\beta-\frac{1}{2}\left (\frac{v}{v_1}\right )^2 \, ,
\end{equation}
with $\zeta=R_c/\sqrt{R_c^2+R_0^2}$, while the velocity $v_1=\sqrt{|\Psi_a|}$ is
fixed through Eq. (\ref{eq:rho_pl}), and with suitable normalizations,
can be cast in the form:
\begin{equation}
v_1=220\;{\rm km\; sec^{-1}} \;\;\left (  \frac{\rho_0}{0.47 \;{\rm
      GeV cm^{-3}} } \right)^{\frac{1}{2}}\frac{R_0}{8.5 \; {\rm kpc}}
~ 
\frac{q (R_c^2+R_0^2)^{(\beta+4)/4}}{\sqrt{|\beta|} R_0
  R_c^{\beta/2} \sqrt{R_c^2 (1
+2 q^2)+R_0^2 (1-\beta q^2)}} \, ,
\end{equation}

The quantities $A$, $B$ and $C$ may be written as:
\begin{eqnarray}
A&=& F_1 \left (\frac{v_{\phi}}{\rm 220 \; km \; sec^{-1}}  \right)^2 
\left (\frac{R_0}{R_c} \right)^4
\frac{\Gamma(2+4/\beta)}{2^{3/2} \pi^{3/2}
  \Gamma(4/\beta-1/2)} \beta (\beta+2) (\frac{1}{q^2}-1) \, ,
\\
B&=& F_1 \left (\frac{v_1}{\rm 220 \; km \; sec^{-1}}  \right)^2 
\left (\frac{R_0}{R_c} \right)^2
\frac{\Gamma(2+4/\beta)}{2^{3/2} \pi^{3/2}
  \Gamma(1/2+4/\beta)} \beta (\beta+2) \frac{1}{q^2} \, , \\
C&=& F_1 \left (\frac{v_1}{\rm 220 \; km \; sec^{-1}}  \right)^2 
\left (\frac{R_0}{R_c} \right)^2
\frac{\Gamma(2+2/\beta)}{2^{3/2} \pi^{3/2}
  \Gamma(1/2+2/\beta)} \beta \left [
  2-\frac{1+\beta}{q^2}\right ] \, , 
\end{eqnarray}
with $F_1=0.47$ GeV cm$^{-3}/ v_1^3$.

For $\beta<0$  Eq. (\ref{eq:dfpl}) still applies, with the following
modifications (now $\Psi_a<0$):
\begin{equation}
\tilde{\epsilon} \equiv \frac{-\epsilon}{\Psi_a}=- \frac{\Psi+\frac{1}{2}
  v^2}{\Psi_a}=\zeta^\beta-\frac{1}{2}\left (\frac{v}{v_1}\right )^2 \, ,
\end{equation}
where:
\begin{eqnarray}
A&=& F_1 \left (\frac{v_{\phi}}{\rm 220 \; km \; sec^{-1}}  \right)^2 
\left (\frac{R_0}{R_c} \right)^4
\frac{\Gamma(3/2-4/\beta)}{2^{3/2} \pi^{3/2}
  \Gamma(-1-4/\beta)} \beta (\beta+2) (1-\frac{1}{q^2}) \, ,
\\
B&=& F_1 \left (\frac{v_1}{\rm 220 \; km \; sec^{-1}}  \right)^2 
\left (\frac{R_0}{R_c} \right)^2
\frac{\Gamma(1/2-4/\beta)}{2^{3/2} \pi^{3/2}
  \Gamma(-1-4/\beta)} \beta (\beta+2) \frac{1}{q^2} \, ,  \\
C&=& F_1 \left (\frac{v_1}{\rm 220 \; km \; sec^{-1}}  \right)^2 
\left (\frac{R_0}{R_c} \right)^2
\frac{\Gamma(1/2-2/\beta)}{2^{3/2} \pi^{3/2}
  \Gamma(-1-2/\beta)} \beta \left [
  \frac{1+\beta}{q^2} -2\right ] \, ,
\end{eqnarray}
with again $F_1=0.47$ GeV cm$^{-3}/ v_1^3$.


\clearpage
\begin{table}
\caption{\label{tab:univ_rho}Values of the parameters for the spherically symmetric density profile of
  Eq.(\ref{eq:univ_rho}).}
\begin{ruledtabular}
\begin{tabular}{|l|cccc|}
                                &  $\alpha$  & $\beta$ & $\gamma$ & $a$ (kpc) \\ 
\hline
NFW \cite{nfw}                  &      1     &    3    &      1   &    20   \\
Moore et al. \cite{moore}       &     1.5    &    3    &     1.5  &    28   \\
Kravtsov et al. \cite{kravtsov} &      2     &    3    &     0.4  &    10   \\
Jaffe \cite{jaffe}              &      1     &    4    &     2    &    160  \\
\end{tabular}
\end{ruledtabular}
\end{table}

\clearpage
\begin{table}
\caption{\label{tab:models}
  Summary of the galactic halo models discussed in Section \ref{sec:models}. 
  The label introduced in the first column is used throughout
  the text to indicate each model in a unique way.
  For all the models
  the numerical values of the parameters which have been used in the
  calculations are given in the third
  column. The last column contains references to the models in the
  text. Models of class C have been analyzed also including
  co--rotation and counter--rotation of the halo through
  Eq. (\ref{eq:corot}). 
}
\begin{ruledtabular}
\begin{tabular}{|c|l|c|c|}
\multicolumn{4}{|l|}{{\bf Class A: Spherical $\bf \rho_{DM}$,  isotropic
  velocity dispersion}} \\ 
\hline
A0 & {\rm ~Isothermal sphere}   &     &     Eq.(\ref{eq:rho_isothermal}) \\
A1 & {\rm ~Evans' logarithmic} \cite{evans_log} &
$R_c=5$ kpc & Eq.(\ref{eq:rho_axi_log_spherical}) \\
A2 & {\rm ~Evans' power--law} \cite{evans_pl} &  
$R_c=16$ kpc, $\beta=0.7$ 
 & Eq.(\ref{eq:rho_pl_spherical}) \\
A3 & {\rm ~Evans' power--law} \cite{evans_pl} & 
$R_c=2$ kpc, $\beta=-0.1$ &
Eq.(\ref{eq:rho_pl_spherical}) \\
A4 & {\rm ~Jaffe} \cite{jaffe}         & Table \ref{tab:univ_rho} & Eq.(\ref{eq:univ_rho}) \\ 
A5 & {\rm ~NFW} \cite{nfw}         & Table \ref{tab:univ_rho} & Eq.(\ref{eq:univ_rho}) \\
A6 & {\rm ~Moore at al.} \cite{moore}   & Table \ref{tab:univ_rho} & Eq.(\ref{eq:univ_rho}) \\ 
A7 & {\rm ~Kravtsov et al.} \cite{kravtsov} & Table \ref{tab:univ_rho} & Eq.(\ref{eq:univ_rho})
\\ 
\hline
\multicolumn{4}{|l|}{{\bf Class B: Spherical $\bf \rho_{DM}$, non--isotropic
  velocity dispersion (Osipkov--Merrit, $\bf \beta_0=0.4$)}} \\ 
\hline 
B1 & {\rm ~Evans' logarithmic} & $R_c=5$ kpc & 
Eqs.(\ref{eq:rho_axi_log_spherical},\ref{eq:beta0}) \\
B2 & {\rm ~Evans' power--law} & 
$R_c=16$ kpc, $\beta=0.7$  & Eqs.(\ref{eq:rho_pl_spherical},\ref{eq:beta0}) \\
B3 & {\rm ~Evans' power--law} & 
$R_c=2$ kpc, $\beta=-0.1$ &
Eqs.(\ref{eq:rho_pl_spherical},\ref{eq:beta0}) \\
B4 & {\rm ~Jaffe}          & Table \ref{tab:univ_rho} &
  Eqs.(\ref{eq:univ_rho},\ref{eq:beta0}) \\ 
B5 & {\rm ~NFW}          & Table \ref{tab:univ_rho} & Eqs.(\ref{eq:univ_rho},\ref{eq:beta0})  \\
B6 & {\rm ~Moore at al.}   & Table \ref{tab:univ_rho} & Eqs.(\ref{eq:univ_rho},\ref{eq:beta0}) \\ 
B7 & {\rm ~Kravtsov et al.} & Table \ref{tab:univ_rho} &
Eqs.(\ref{eq:univ_rho},\ref{eq:beta0}) \\
\hline
\multicolumn{4}{|l|}{{\bf Class C: Axisymmetric $\bf \rho_{DM}$}}  \\ 
\hline
C1 & {\rm ~Evans' logarithmic} & $R_c=0$, $q=1/\sqrt{2}$ &
Eqs.(\ref{eq:potential_axi_log},\ref{eq:rho_axi_log}) \\
C2 & {\rm ~Evans' logarithmic} &
$R_c=5$ kpc, $q=1/\sqrt{2}$ &
Eqs.(\ref{eq:potential_axi_log},\ref{eq:rho_axi_log}) \\
C3 & {\rm ~Evans' power--law} & 
$R_c=16$ kpc, $q=0.95$, $\beta=0.9$  & Eqs.(\ref{eq:potential_pl},\ref{eq:rho_pl}) \\
C4 & {\rm ~Evans' power--law}& 
$R_c=2$ kpc, $q=1/\sqrt{2}$, $\beta=-0.1$ &
Eqs.(\ref{eq:potential_pl},\ref{eq:rho_pl}) \\
\hline
\multicolumn{4}{|l|}{{\bf Class D: Triaxial $\bf \rho_{DM}$ \cite{triaxial}
  ($\bf q=0.8$, $\bf p=0.9$)}}  \\ 
\hline
D1 & {\rm ~Earth on major axis, radial anisotropy}  & $\delta=-1.78$  &
Eqs.(\ref{eq:potential_triaxial},\ref{eq:rho_triaxial}) \\
D2 & {\rm ~Earth on major axis, tangential anis.} &   $\delta=16$ &
Eqs.(\ref{eq:potential_triaxial},\ref{eq:rho_triaxial}) \\
D3 & {\rm ~Earth on intermediate axis, radial anis.}  &  $\delta=-1.78$ &
Eqs.(\ref{eq:potential_triaxial},\ref{eq:rho_triaxial}) \\
D4 & {\rm ~Earth on intermediate axis, tangential anis.} & $\delta=16$ &
Eqs.(\ref{eq:potential_triaxial},\ref{eq:rho_triaxial}) \\
\end{tabular}
\end{ruledtabular}
\end{table}

\clearpage
\begin{table}
\caption{\label{tab:intervals} Allowed intervals of $\rho_0$ obtained from the
  constraints on $M_{vis}$ and $v_{rot}^{100}$ for the halo models summarized
  in Table \ref{tab:models}. The values of $\rho_0^{max}$
  and $\rho_0^{min}$
  are used in the modulation analysis of the experimental data of
  Figs. \ref{fig:a0}--\ref{fig:b7} for the models of class A and B, while only $\rho_0^{max}$
  is used for models of class C and D in Figs. \ref{fig:c1}--\ref{fig:d4}. The value of $\rho_0^{max}$ for the
  axisymmetric models of class C are not affected by the inclusion 
  of a co--rotation or
  counter--rotation effect through Eq.(\ref{eq:corot}).
}
\begin{ruledtabular}
\begin{tabular}{|c|c|c|c|c|c|c|}
& \multicolumn{2}{c|} {$v_0 =170$ km sec$^{-1}$} & \multicolumn{2}{c|} {$v_0 =220$ km sec$^{-1}$}
& \multicolumn{2}{c|} {$v_0 =270$ km sec$^{-1}$} \\ 
\hline
Model &  $\rho_0^{min}$  & $\rho_0^{max}$ &  $\rho_0^{min}$  & $\rho_0^{max}$ &  $\rho_0^{min}$  & $\rho_0^{max}$ 
\\ 
\hline 
  A0           & 0.18 & 0.28 & 0.30 & 0.47  & 0.45 & 0.71 \\
  A1 ,  B1     & 0.20 & 0.42 & 0.34 & 0.71  & 0.62 & 1.07 \\
  A2 ,  B2     & 0.24 & 0.53 & 0.41 & 0.89  & 0.97 & 1.33 \\
  A3 ,  B3     & 0.17 & 0.35 & 0.29 & 0.59  & 0.52 & 0.88 \\
  A4 ,  B4     & 0.26 & 0.27 & 0.44 & 0.45  & 0.66 & 0.67 \\
  A5 ,  B5     & 0.20 & 0.44 & 0.33 & 0.74  & 0.66 & 1.11 \\
  A6 ,  B6     & 0.22 & 0.39 & 0.37 & 0.65  & 0.57 & 0.98 \\
  A7 ,  B7     & 0.32 & 0.54 & 0.54 & 0.91  & 0.82 & 1.37 \\
  C1           & 0.36 & 0.56 & 0.60 & 0.94  & 0.91 & 1.42 \\
  C2           & 0.34 & 0.67 & 0.56 & 1.11  & 0.98 & 1.68 \\
  C3           & 0.30 & 0.66 & 0.50 & 1.10  & 0.97 & 1.66 \\
  C4           & 0.32 & 0.65 & 0.54 & 1.09  & 0.96 & 1.64 \\
  D1 ,  D2     & 0.32 & 0.50 & 0.54 & 0.84  & 0.81 & 1.27 \\
  D3 ,  D4     & 0.19 & 0.30 & 0.32 & 0.51  & 0.49 & 0.76
\end{tabular}
\end{ruledtabular}
\end{table}


\clearpage

\clearpage

\clearpage

\clearpage

\begin{figure}\centering
  \rotatebox{270}{\includegraphics[bb=195 85 435 720,width=0.34\textwidth]{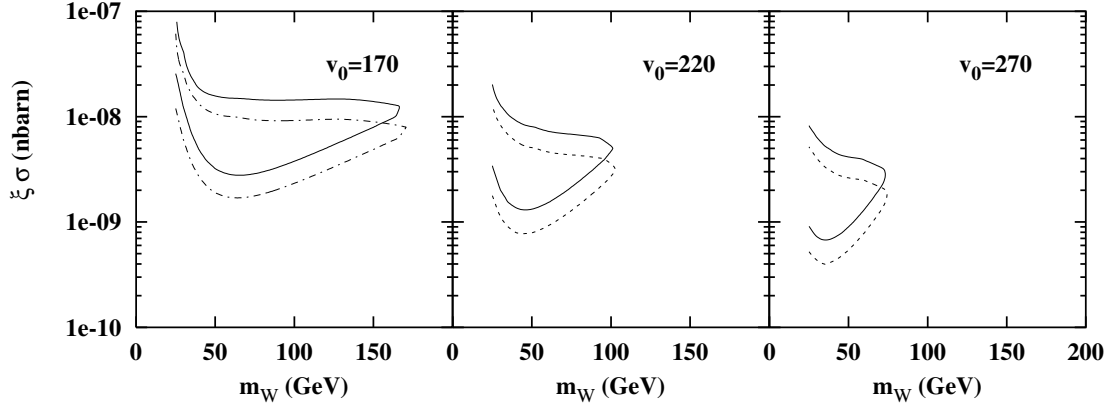}} 
  \caption{
    Plot of the 3$\sigma$ annual--modulation region in the plane $\xi
    \sigma_{\rm scalar}^{\rm (nucleon)}$ versus $m_{W}$ using for the velocity
    distribution of WIMPs the isothermal sphere model (Model A0, see Table
    \ref{tab:models}).  The three panels of the figure correspond to $v_0$=170,
    220, 270 km sec$^{-1}$ from left to right. Upper (lower) regions correspond
    to $\rho_0$=$\rho_0^{min}$ ($\rho_0^{max}$) where $\rho_0^{min}$ and
    $\rho_0^{max}$ are given in Table \ref{tab:intervals}.
\label{fig:a0}}
\end{figure}

\begin{figure}\centering
  \rotatebox{270}{\includegraphics[bb=195 85 435 720,width=0.34\textwidth]{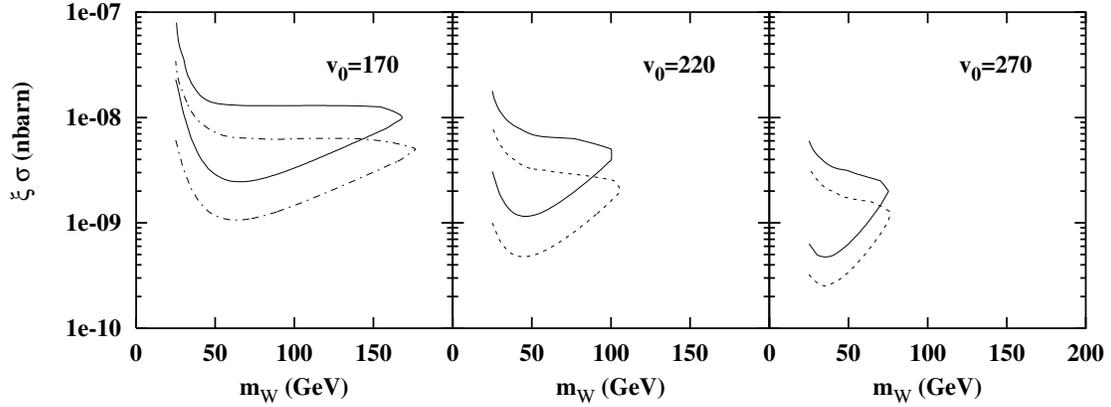}}
  \caption{
    The same as in Fig. \ref{fig:a0} for the velocity distribution of
    model A1 (see table \ref{tab:models}).
\label{fig:a1}}
\end{figure}

\begin{figure}\centering
  \rotatebox{270}{\includegraphics[bb=195 85 435 720,width=0.34\textwidth]{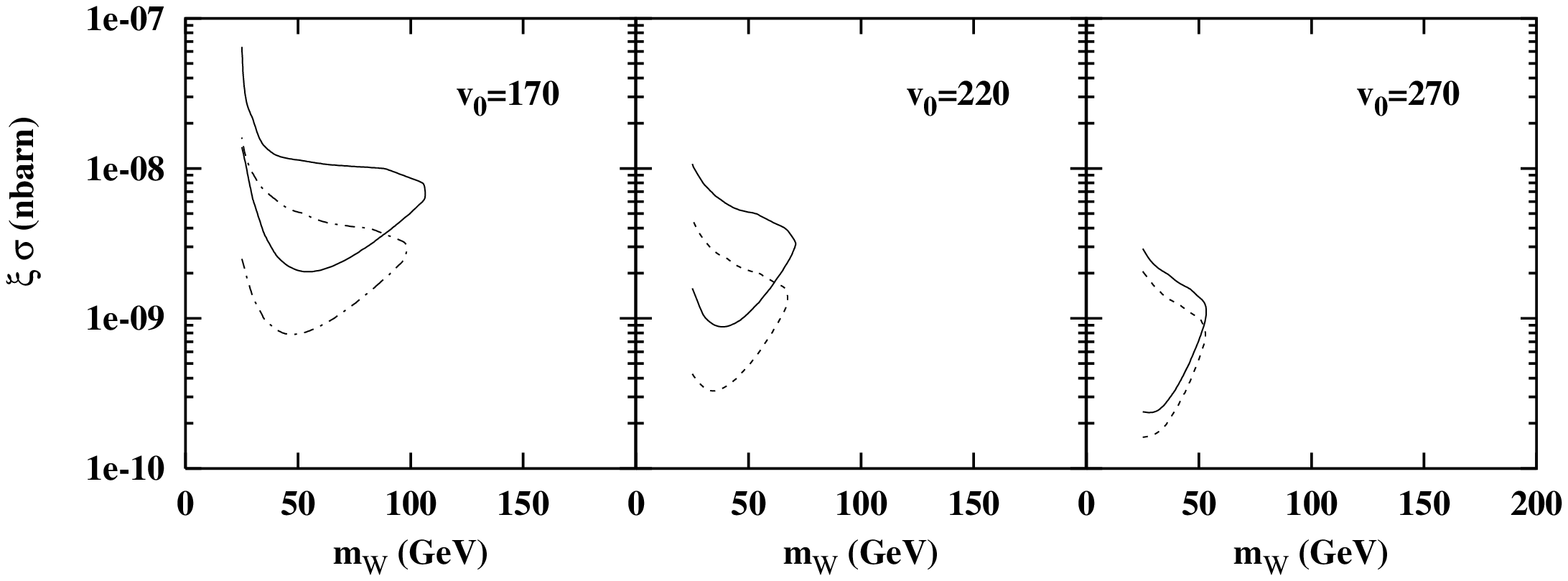}}
  \caption{
    The same as in Fig. \ref{fig:a0} for the velocity distribution of
    model A2.
\label{fig:a2}}
\end{figure}

\begin{figure}\centering
  \rotatebox{270}{\includegraphics[bb=195 85 435 720,width=0.34\textwidth]{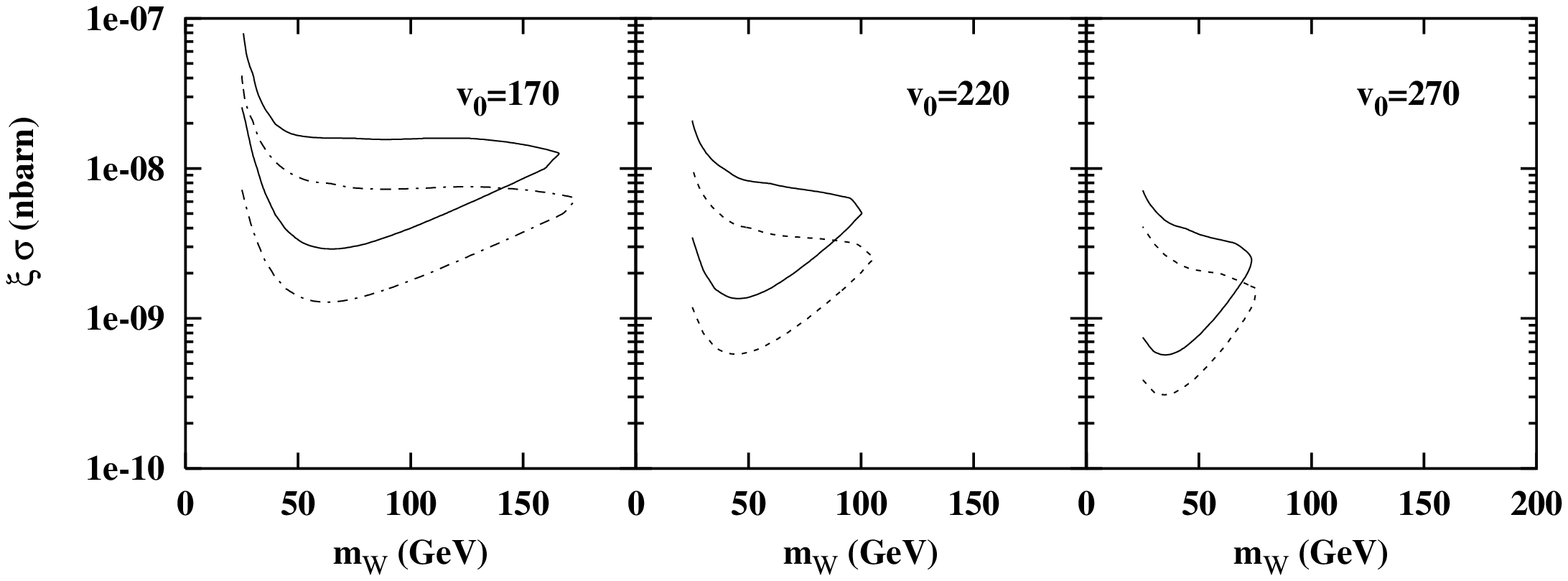}}
  \caption{
    The same as in Fig. \ref{fig:a0} for the velocity distribution of
    model A3.
\label{fig:a3}}
\end{figure}

\begin{figure}\centering
  \rotatebox{270}{\includegraphics[bb=195 85 435 720,width=0.34\textwidth]{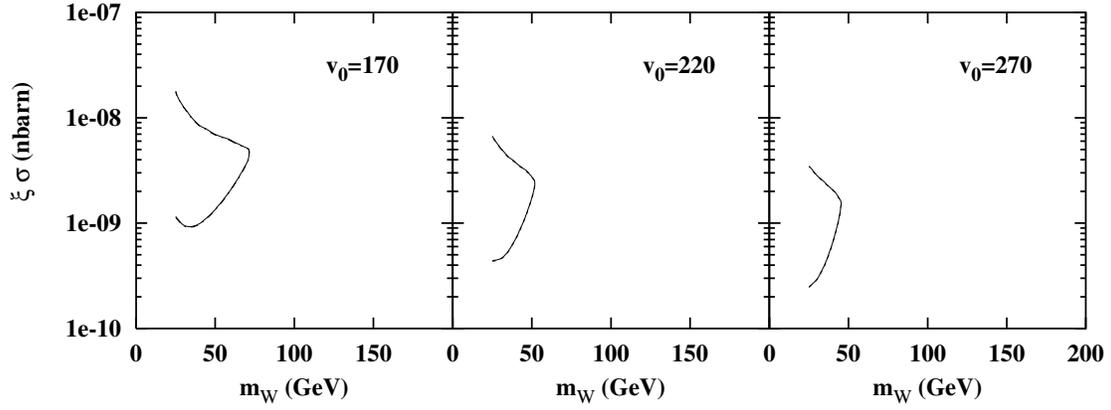}}
  \caption{
    The same as in Fig. \ref{fig:a0} for the velocity distribution of
    model A4. In this case $\rho_0^{max}\simeq\rho_0^{min}$, so upper
    and lower curves are not distinguishable.
\label{fig:a4}}
\end{figure}

\begin{figure}\centering
  \rotatebox{270}{\includegraphics[bb=195 85 435 720,width=0.34\textwidth]{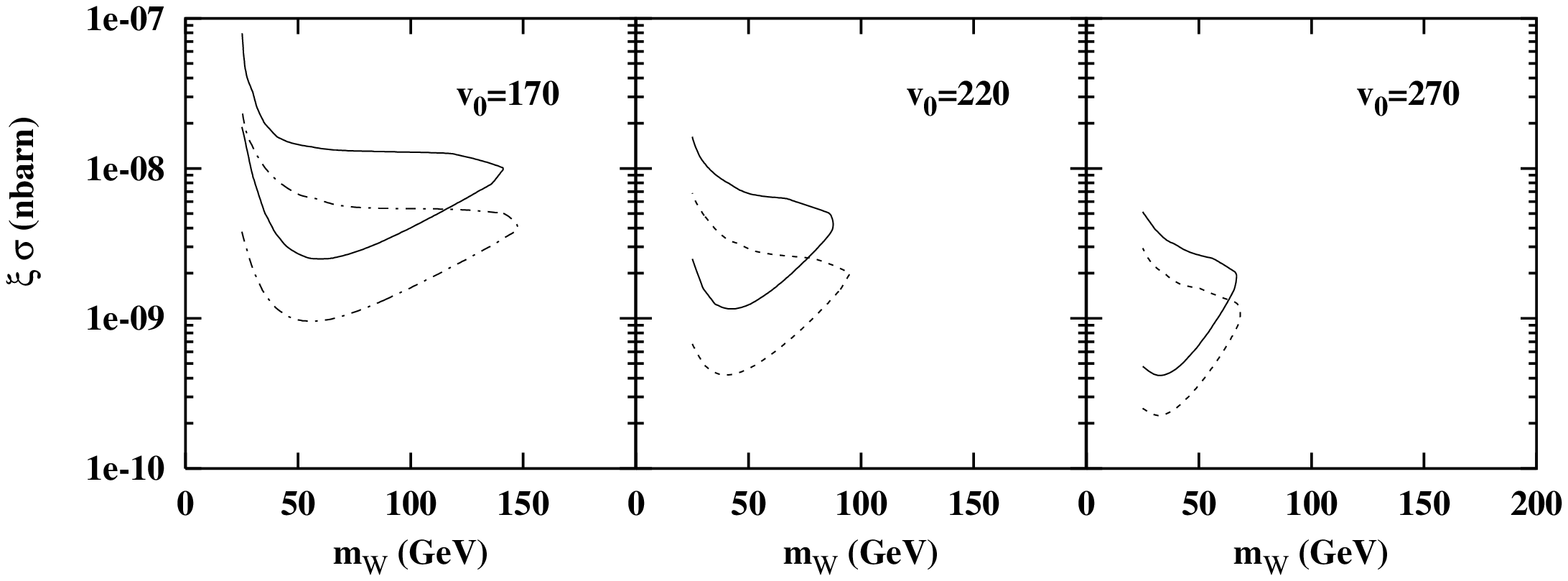}}
  \caption{
    The same as in Fig. \ref{fig:a0} for the velocity distribution of
    model A5.
\label{fig:a5}}
\end{figure}

\begin{figure}\centering
  \rotatebox{270}{\includegraphics[bb=195 85 435 720,width=0.34\textwidth]{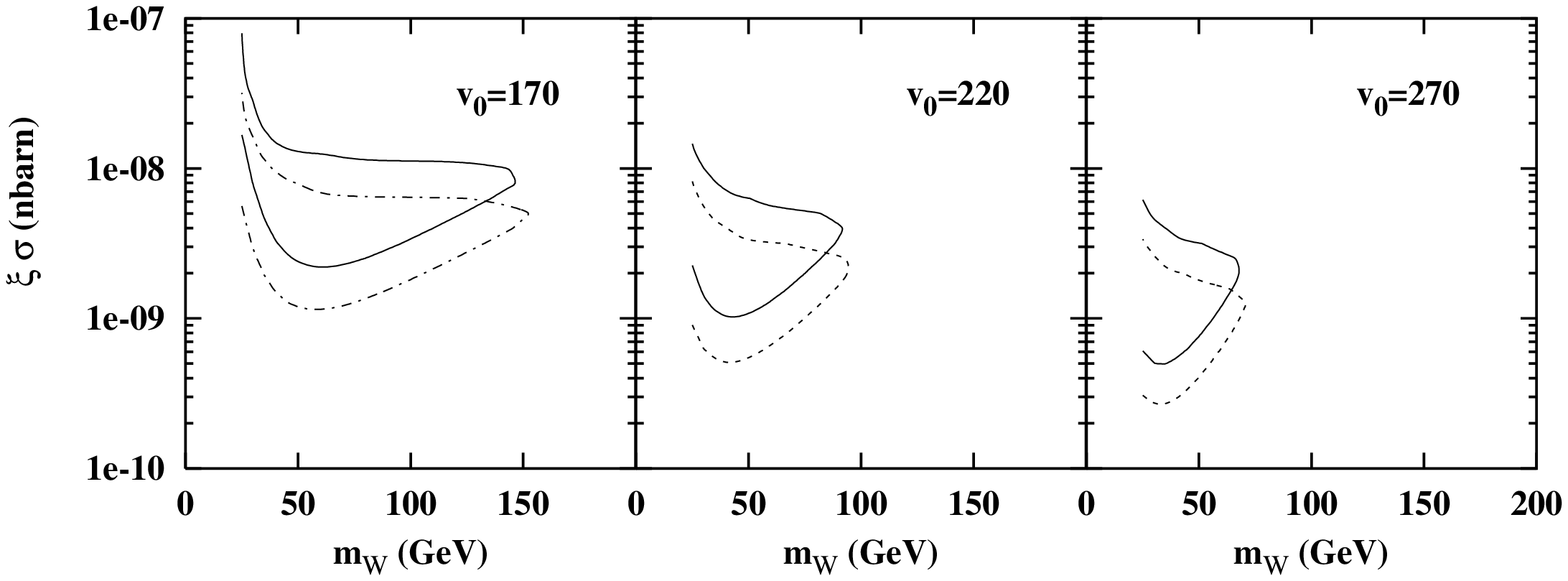}}
  \caption{
    The same as in Fig. \ref{fig:a0} for the velocity distribution of
    model A6.
\label{fig:a6}}
\end{figure}

\begin{figure}\centering
  \rotatebox{270}{\includegraphics[bb=195 85 435 720,width=0.34\textwidth]{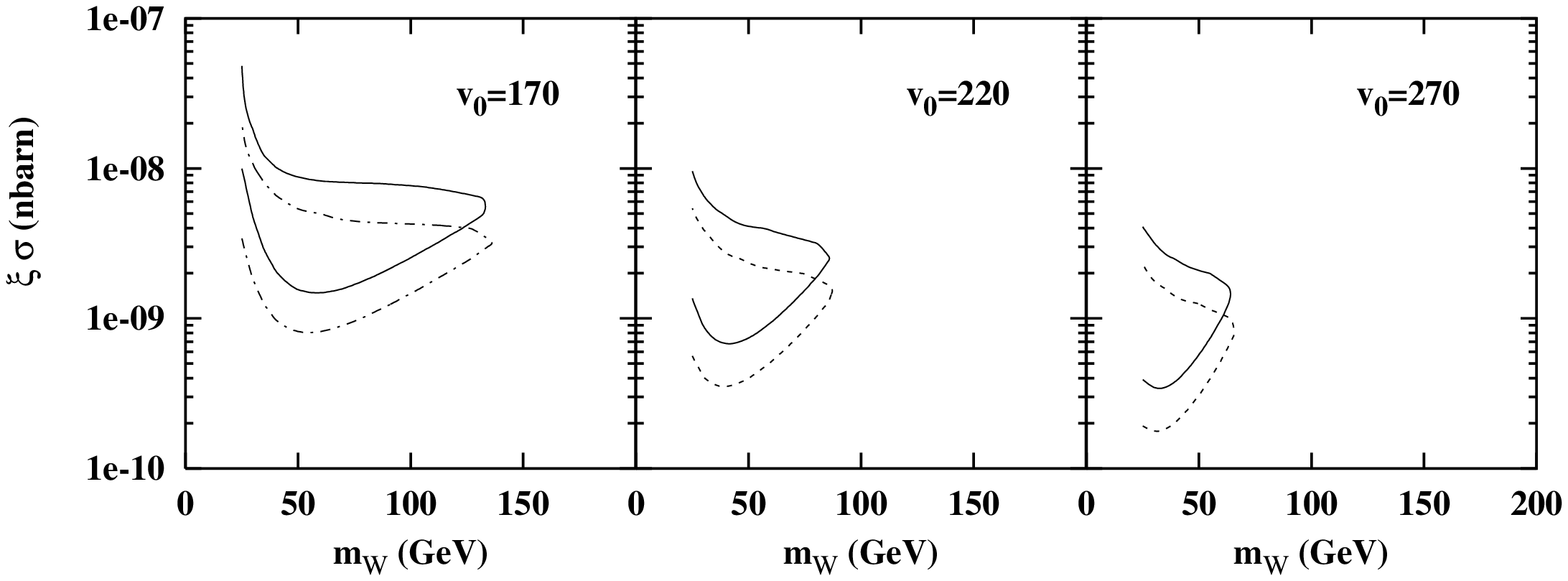}}
  \caption{
    The same as in Fig. \ref{fig:a0} for the velocity distribution of
    model A7.
\label{fig:a7}}
\end{figure}

\begin{figure}\centering
  \rotatebox{270}{\includegraphics[bb=195 85 435 720,width=0.34\textwidth]{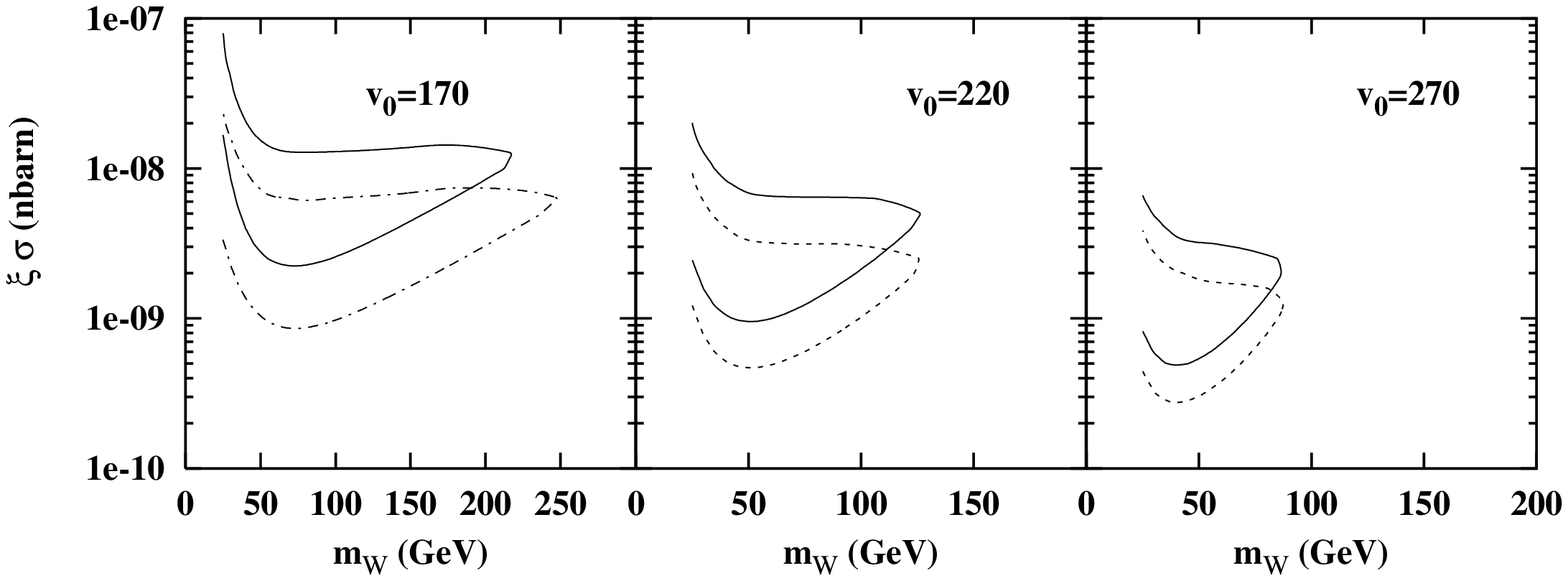}}
  \caption{
    The same as in Fig. \ref{fig:a1}, with anisotropy of the velocity
    dispersion through the Osipkov--Merrit term and anisotropy parameter
    $\beta_0=0.4$ (model B1).  The horizontal axis has been extended in the
    first panel.
\label{fig:b1}}
\end{figure}

\begin{figure}\centering
  \rotatebox{270}{\includegraphics[bb=195 85 435 720,width=0.34\textwidth]{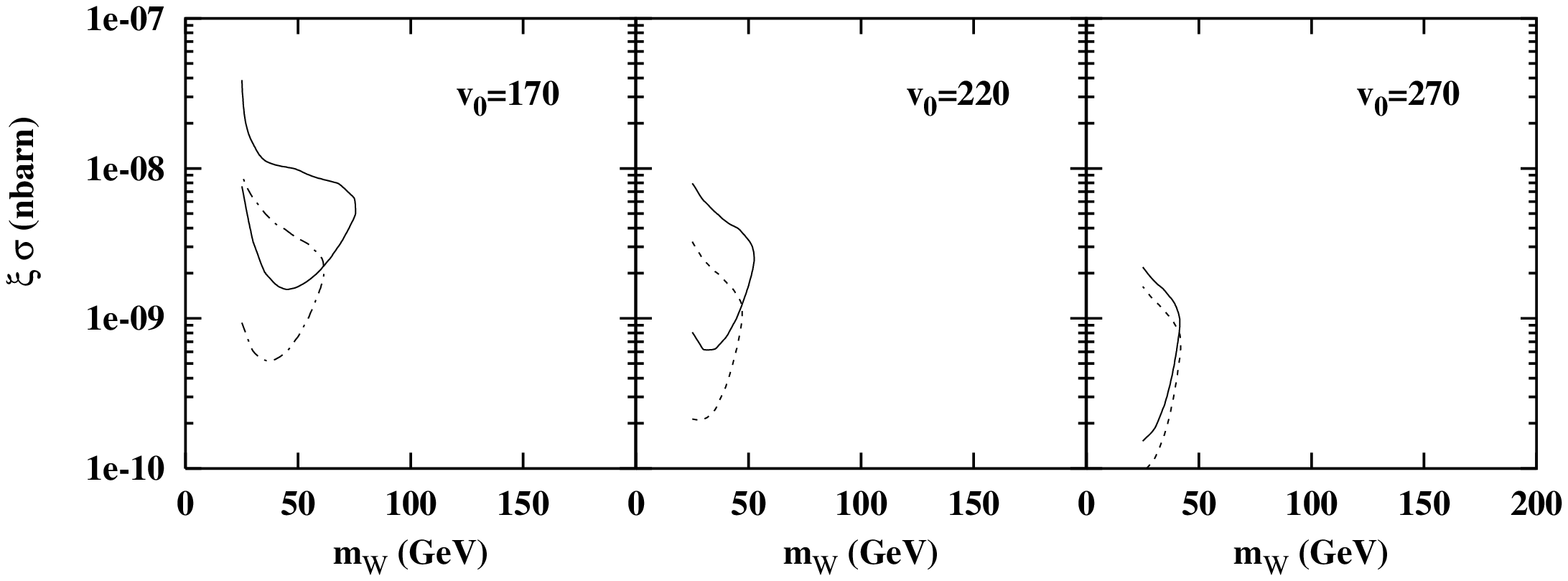}}
  \caption{
    The same as in Fig. \ref{fig:a2}, with anisotropy of the velocity
    dispersion through the Osipkov--Merrit term and anisotropy parameter
    $\beta_0=0.4$ (model B2).
\label{fig:b2}}
\end{figure}

\begin{figure}\centering
  \rotatebox{270}{\includegraphics[bb=195 85 435 720,width=0.34\textwidth]{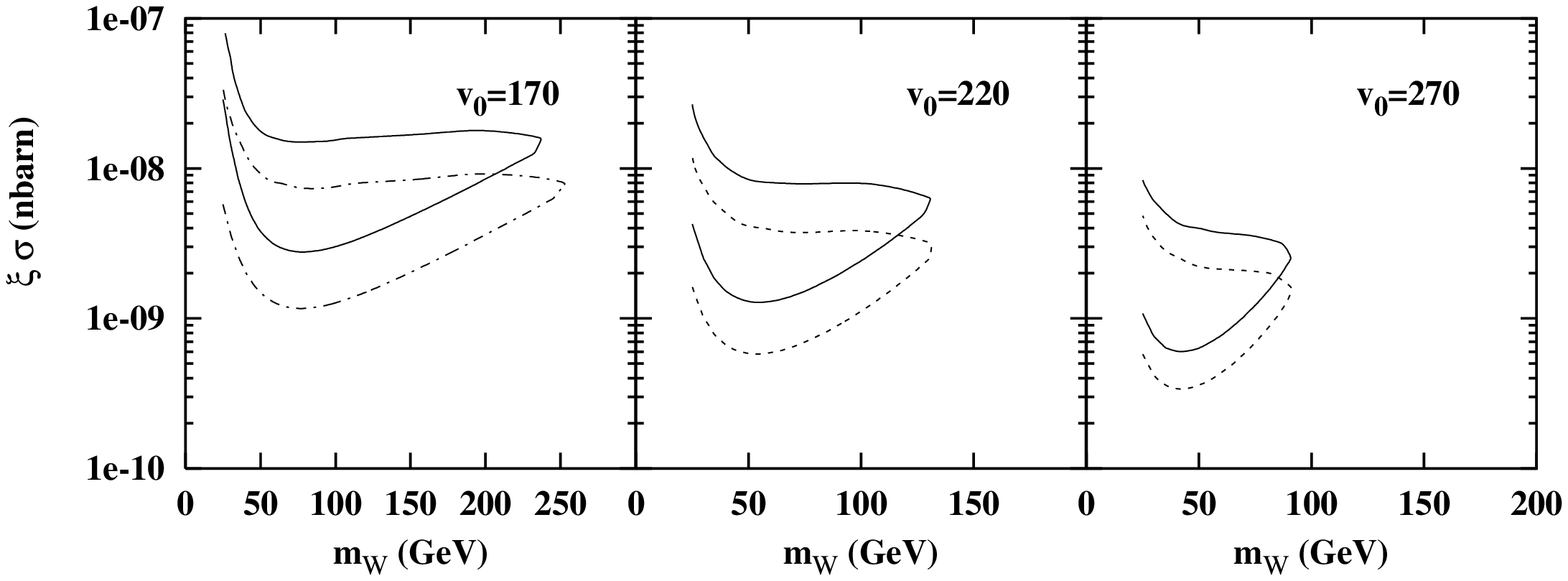}}
  \caption{
    The same as in Fig. \ref{fig:a3}, with anisotropy of the velocity
    dispersion through the Osipkov--Merrit term and anisotropy parameter
    $\beta_0=0.4$ (model B3). The
    horizontal axis has been extended in the first panel.
\label{fig:b3}}
\end{figure}

\begin{figure}\centering
  \rotatebox{270}{\includegraphics[bb=195 85 435 720,width=0.34\textwidth]{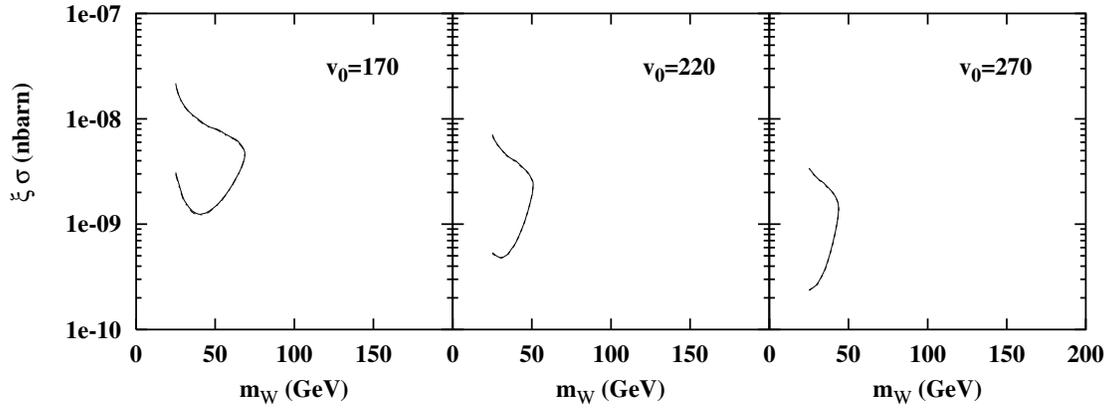}}
  \caption{
    The same as in Fig. \ref{fig:a4}, with anisotropy of the velocity
    dispersion through the Osipkov--Merrit term and anisotropy parameter
    $\beta_0=0.4$ (model B4). In this case $\rho_0^{max}\simeq\rho_0^{min}$, so
    upper and lower curves are not distinguishable.
\label{fig:b4}}
\end{figure}

\begin{figure}\centering
  \rotatebox{270}{\includegraphics[bb=195 85 435 720,width=0.34\textwidth]{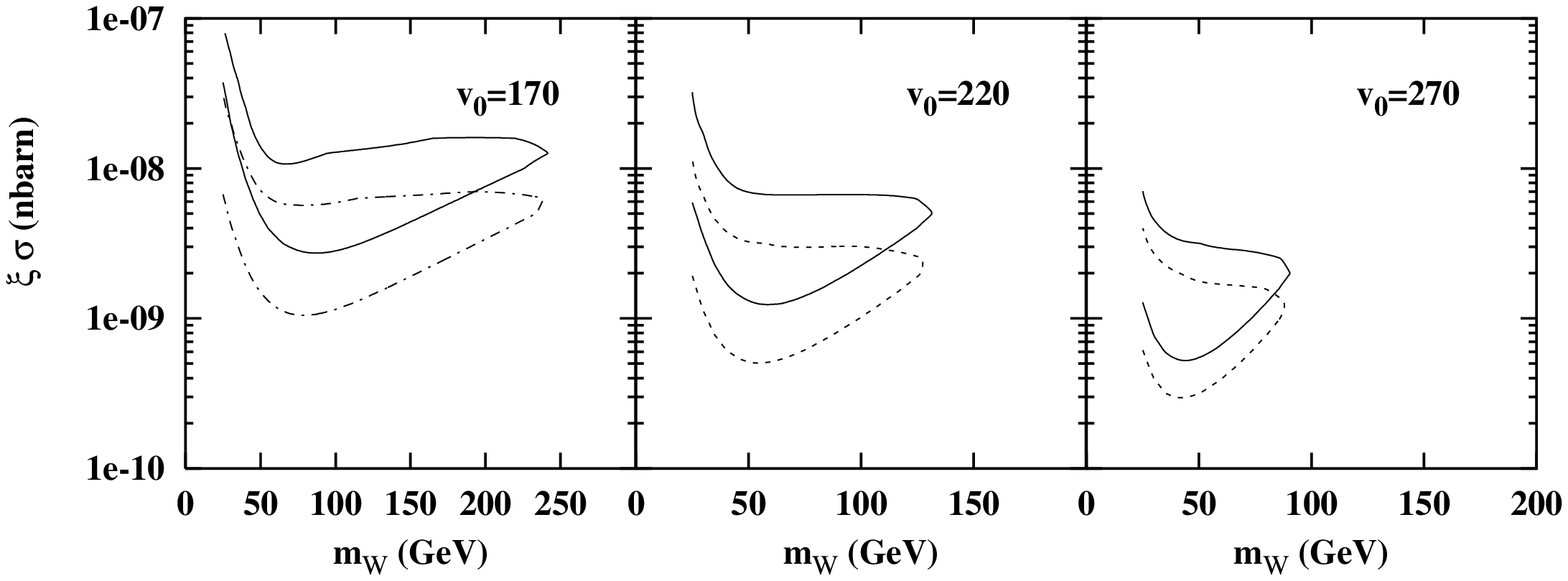}}
  \caption{
    The same as in Fig. \ref{fig:a5}, with anisotropy of the velocity
    dispersion through the Osipkov--Merrit term and anisotropy parameter
    $\beta_0=0.4$ (model B5). The
    horizontal axis has been extended in the first panel.
\label{fig:b5}}
\end{figure}

\begin{figure}\centering
  \rotatebox{270}{\includegraphics[bb=195 85 435 720,width=0.34\textwidth]{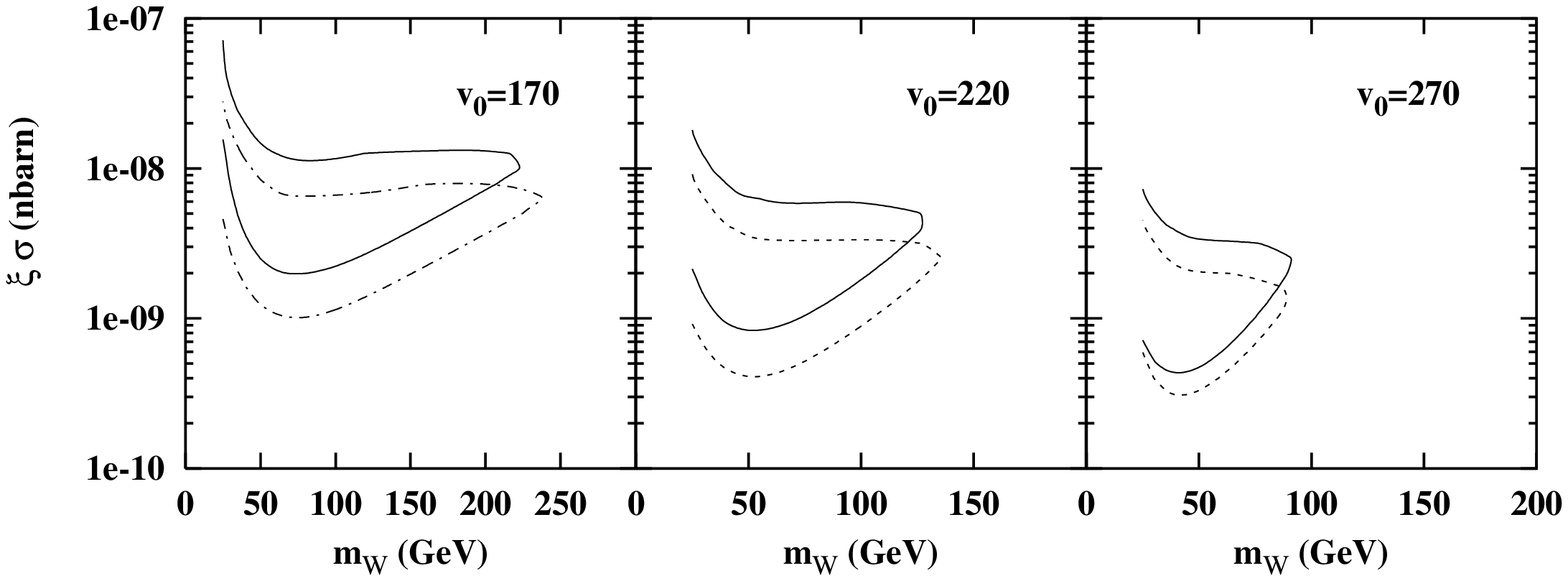}}
  \caption{
    The same as in Fig. \ref{fig:a6}, with anisotropy of the velocity
    dispersion through the Osipkov--Merrit term and anisotropy parameter
    $\beta_0=0.4$ (model B6). The
    horizontal axis has been extended in the first panel.
\label{fig:b6}}
\end{figure}

\begin{figure}\centering
  \rotatebox{270}{\includegraphics[bb=195 85 435 720,width=0.34\textwidth]{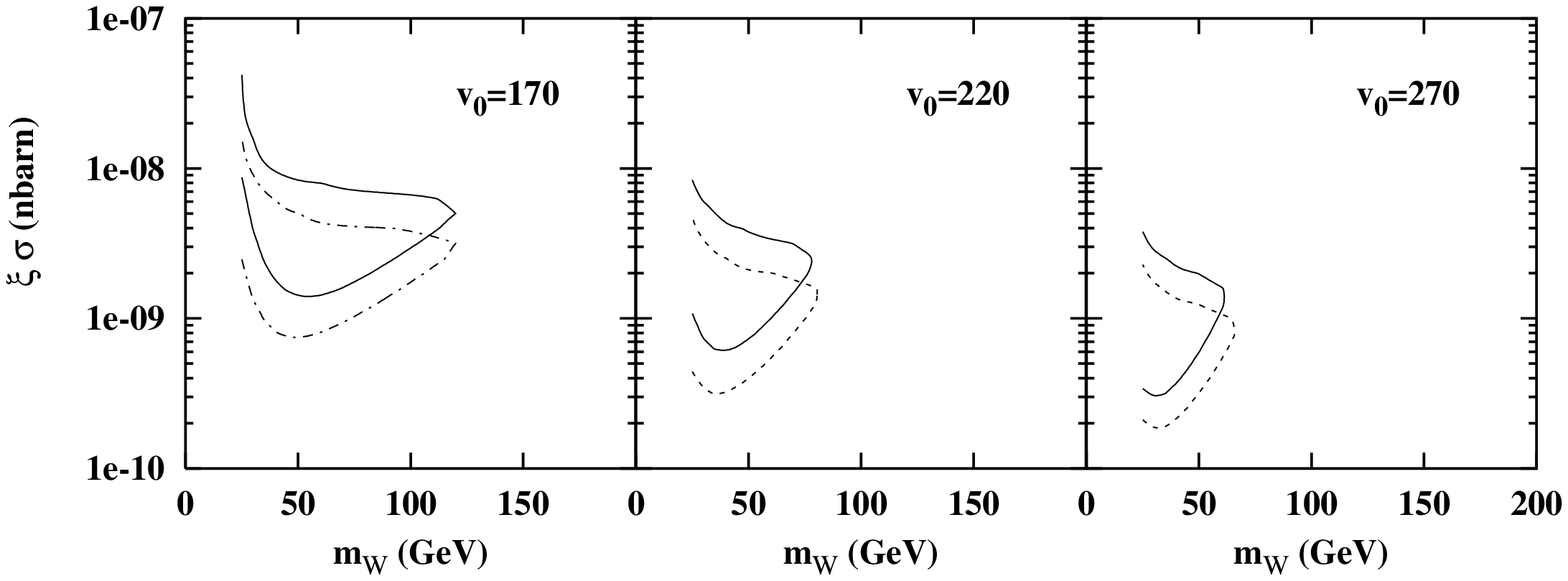}}
  \caption{
    The same as in Fig. \ref{fig:a7}, with anisotropy of the velocity
    dispersion through the Osipkov--Merrit term and anisotropy parameter
    $\beta_0=0.4$ (model B7).
\label{fig:b7}}
\end{figure}

\clearpage
 
\begin{figure}\centering
  \rotatebox{270}{\includegraphics[bb=195 85 435 720,width=0.34\textwidth]{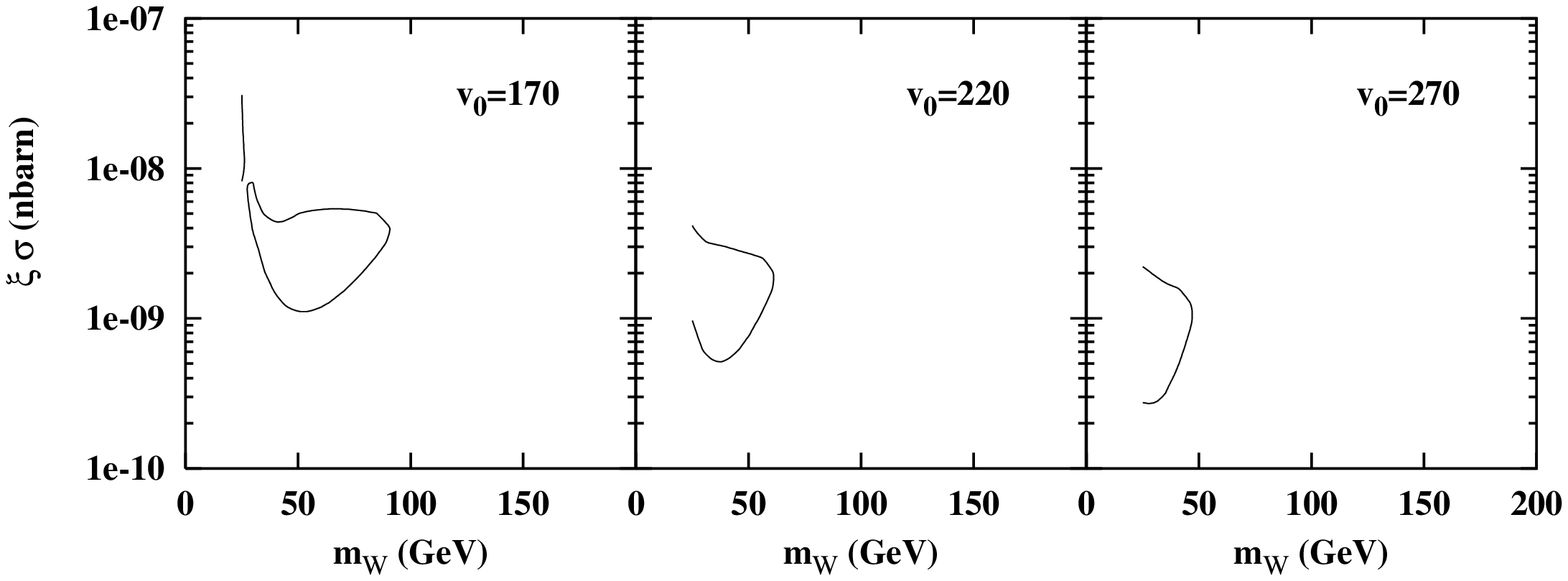}}
  \caption{
    The same as in Fig. \ref{fig:a0} for the velocity distribution of model C1.
    Only the case $\rho_0$=$\rho_0^{max}$ is shown.
\label{fig:c1}}
\end{figure}

\begin{figure}\centering
  \rotatebox{270}{\includegraphics[bb=195 85 435 720,width=0.34\textwidth]{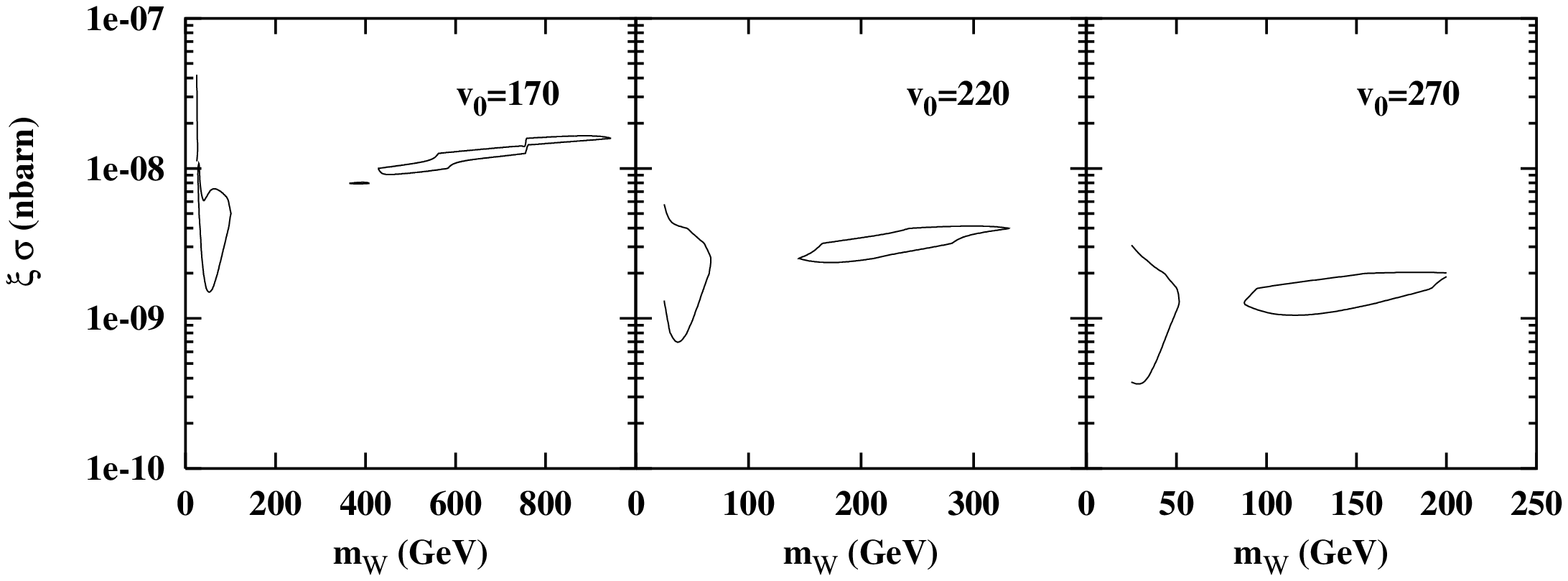}}
  \caption{
    The same as in Fig. \ref{fig:c1} including a co--rotation effect of the
    halo with $\eta=0.64$. The horizontal axis has been extended in all panels.
\label{fig:c1_corot}}
\end{figure}

\begin{figure}\centering
  \rotatebox{270}{\includegraphics[bb=195 85 435 720,width=0.34\textwidth]{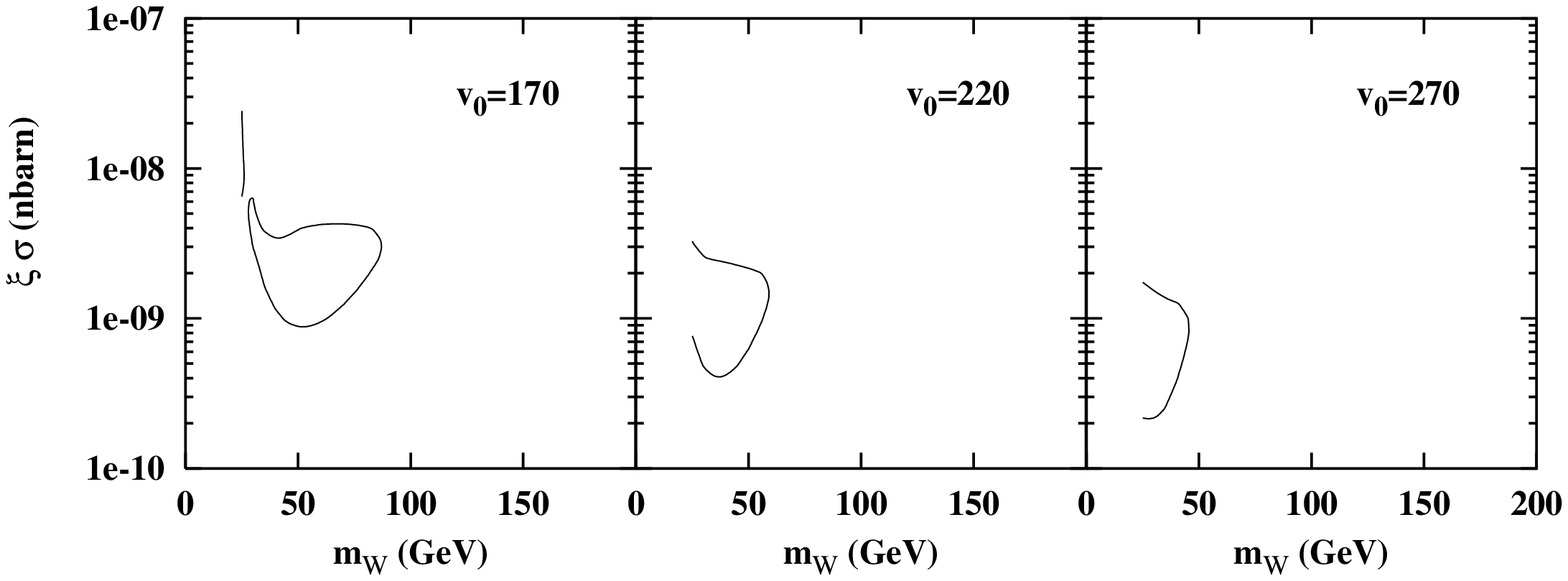}}
  \caption{
    The same as in Fig. \ref{fig:c1} including a counter--rotation effect of
    the halo with $\eta=0.36$.
\label{fig:c1_controrot}}
\end{figure}

\begin{figure}\centering
  \rotatebox{270}{\includegraphics[bb=195 85 435 720,width=0.34\textwidth]{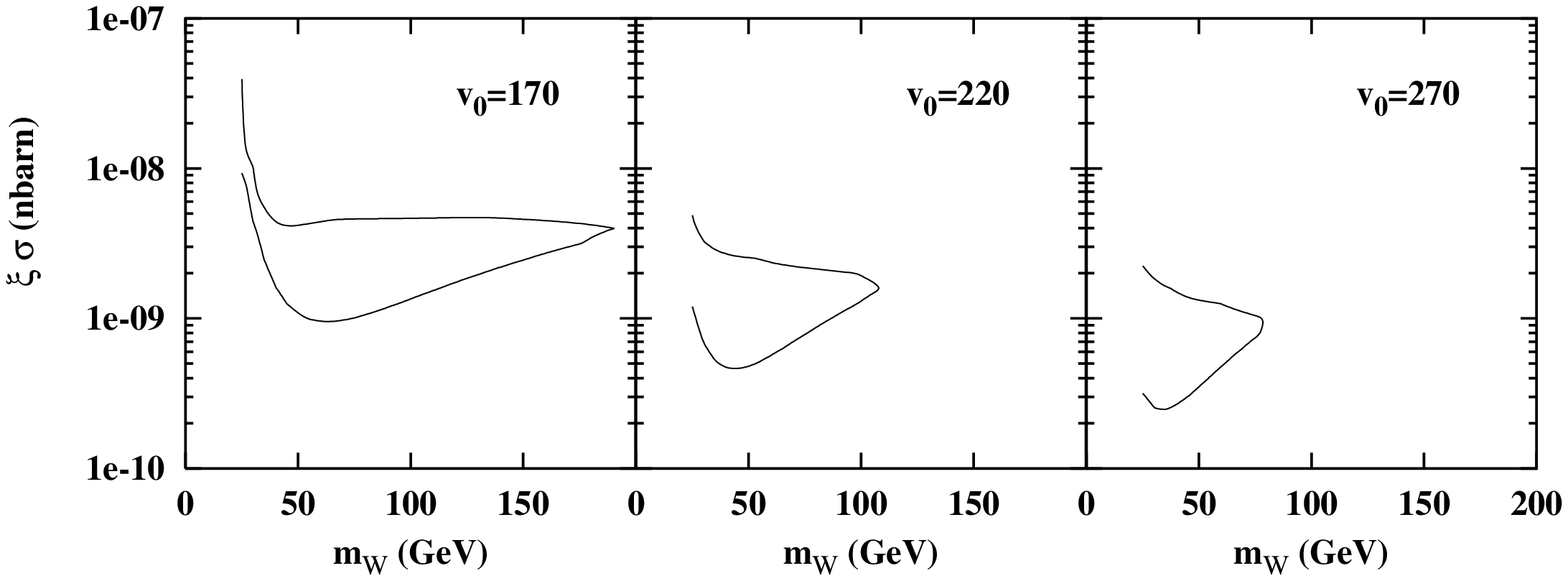}}
  \caption{
    The same as in Fig. \ref{fig:a0} for the velocity distribution of
    model C2. Only the case $\rho_0$=$\rho_0^{max}$
    is shown.
\label{fig:c2}}
\end{figure}

\begin{figure}\centering
  \rotatebox{270}{\includegraphics[bb=195 85 435 720,width=0.34\textwidth]{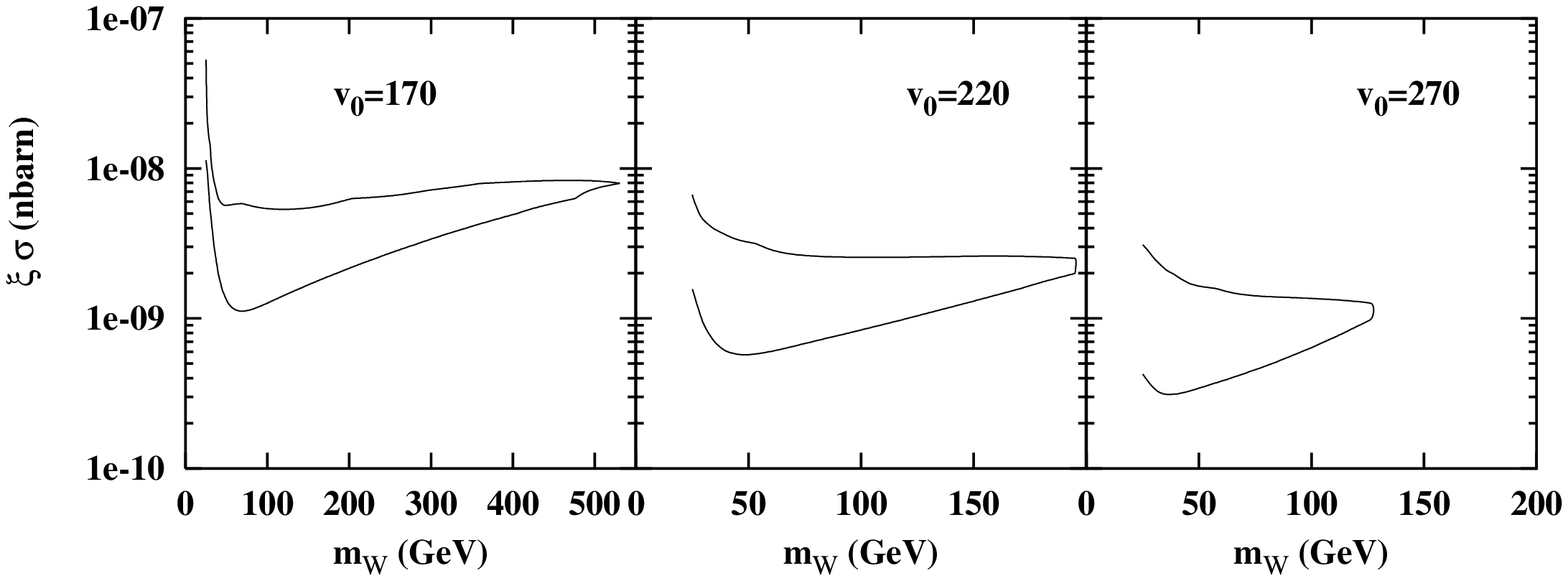}}
  \caption{
    The same as in Fig. \ref{fig:c2} including a co--rotation effect of the halo
    with $\eta=0.64$. The horizontal axis has been
    extended in the first panel.
\label{fig:c2_corot}}
\end{figure}

\begin{figure}\centering
  \rotatebox{270}{\includegraphics[bb=195 85 435 720,width=0.34\textwidth]{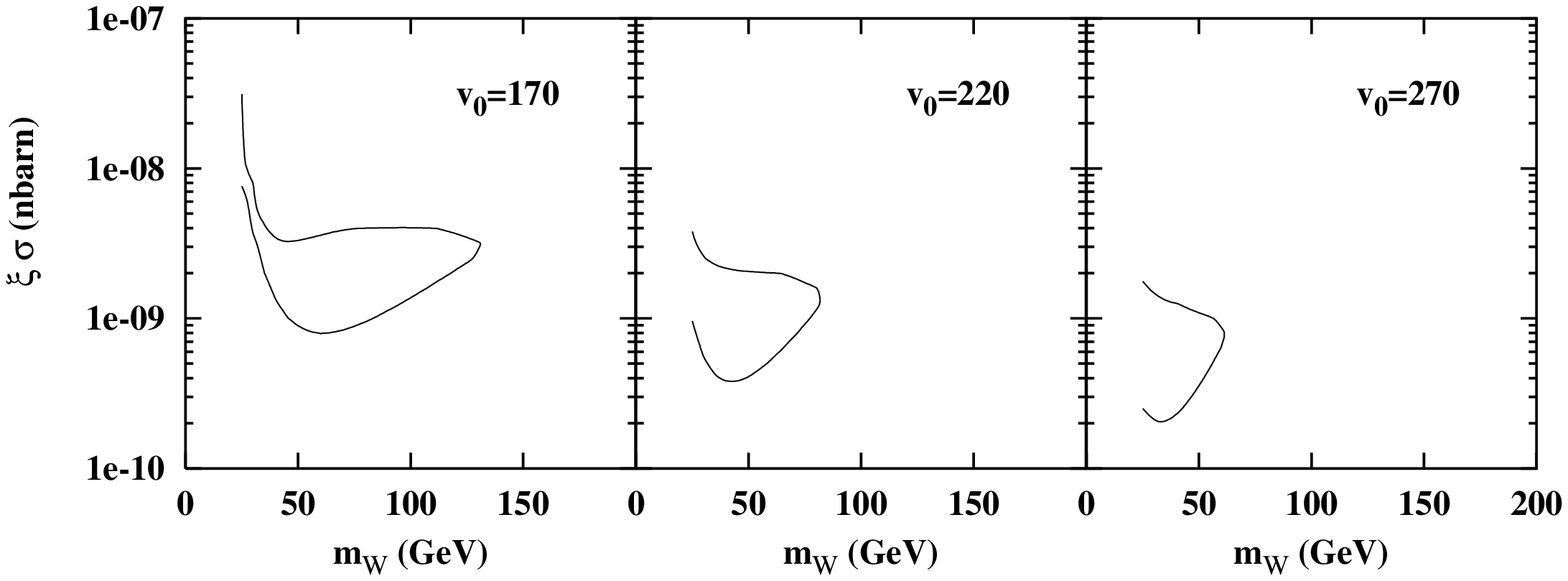}}
  \caption{
    The same as in Fig. \ref{fig:c2} including a counter--rotation effect of
    the halo with $\eta=0.36$.
\label{fig:c2_controrot}}
\end{figure}

\begin{figure}\centering
  \rotatebox{270}{\includegraphics[bb=195 85 435 720,width=0.34\textwidth]{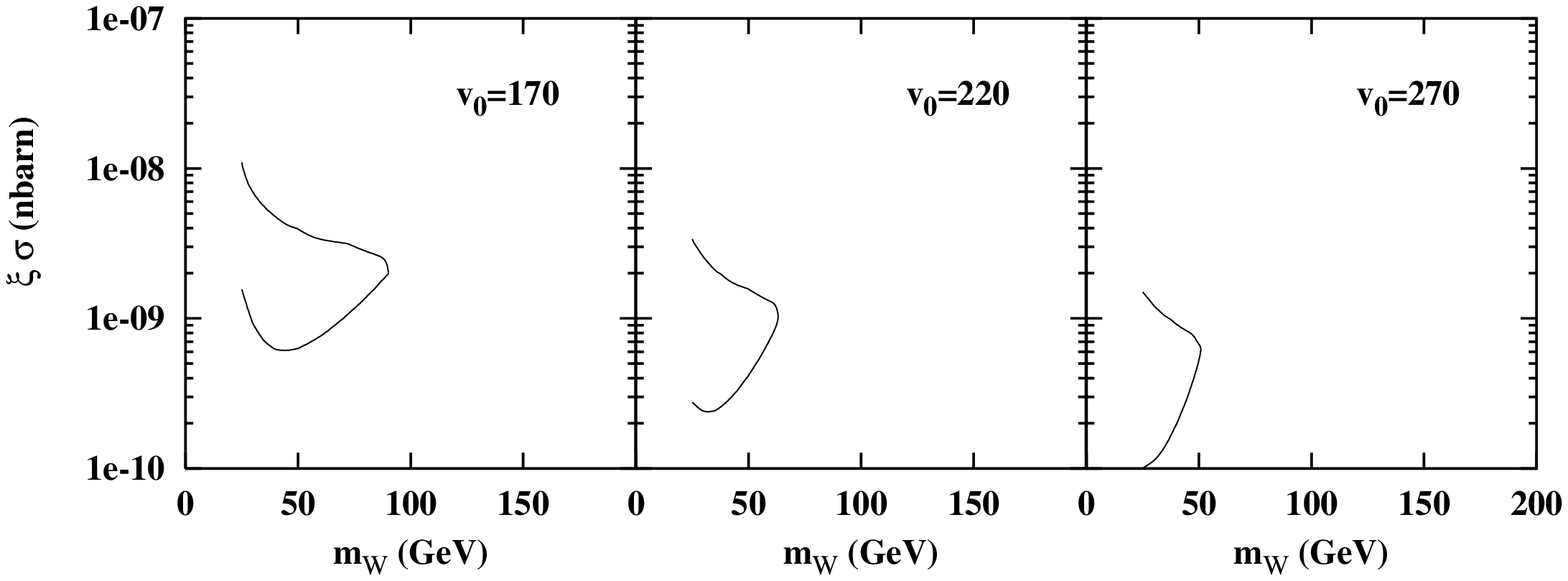}}
  \caption{
    The same as in Fig. \ref{fig:a0} for the velocity distribution of model C3.
    Only the case $\rho_0$=$\rho_0^{max}$ is shown.
\label{fig:c3}}
\end{figure}

\begin{figure}\centering
  \rotatebox{270}{\includegraphics[bb=195 85 435 720,width=0.34\textwidth]{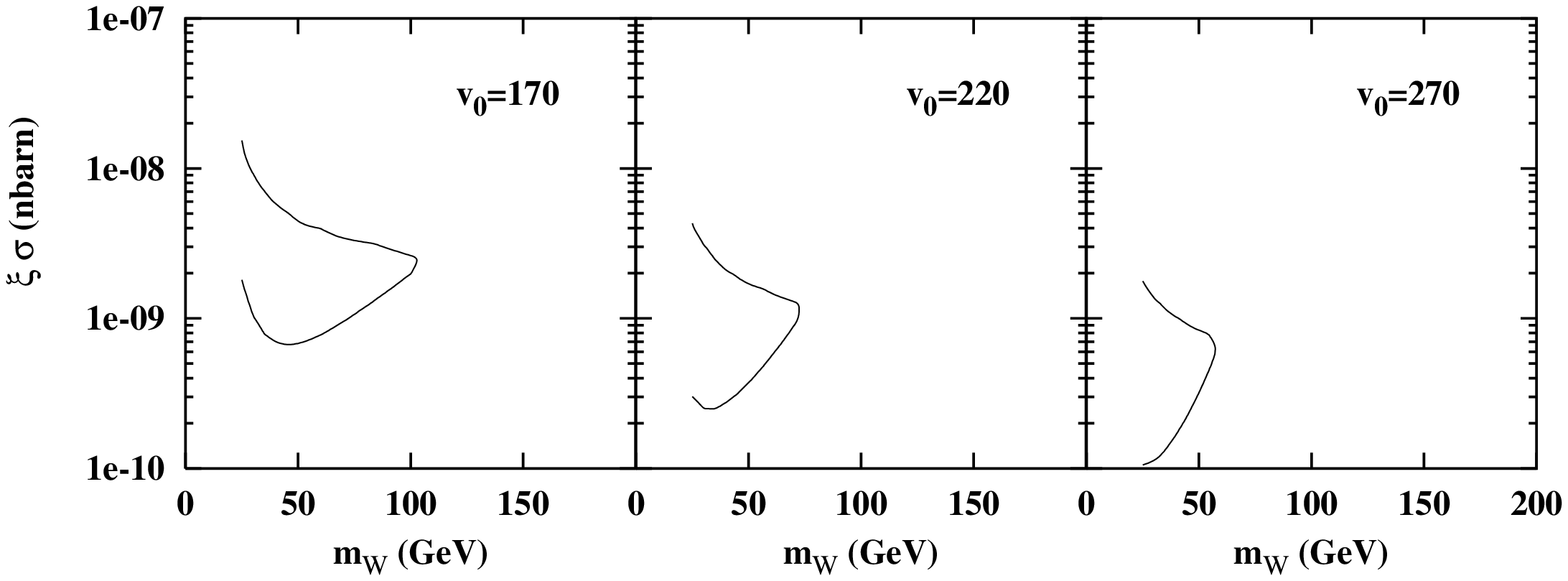}}
  \caption{
    The same as in Fig. \ref{fig:c3} including a co--rotation effect of the halo
    with $\eta=0.64$.
\label{fig:c3_corot}}
\end{figure}

\begin{figure}\centering
  \rotatebox{270}{\includegraphics[bb=195 85 435 720,width=0.34\textwidth]{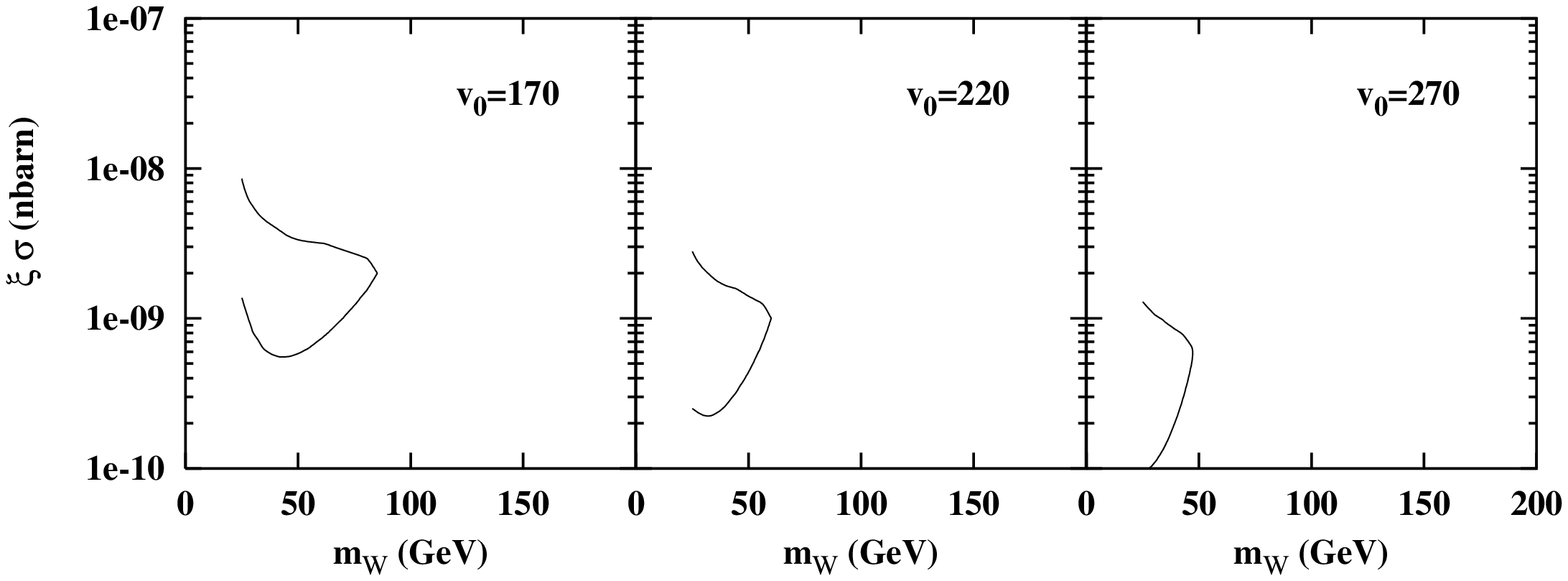}}
  \caption{
    The same as in Fig. \ref{fig:c3} including a counter--rotation effect of
    the halo with $\eta=0.36$.
\label{fig:c3_controrot}}
\end{figure}

\begin{figure}\centering
  \rotatebox{270}{\includegraphics[bb=195 85 435 720,width=0.34\textwidth]{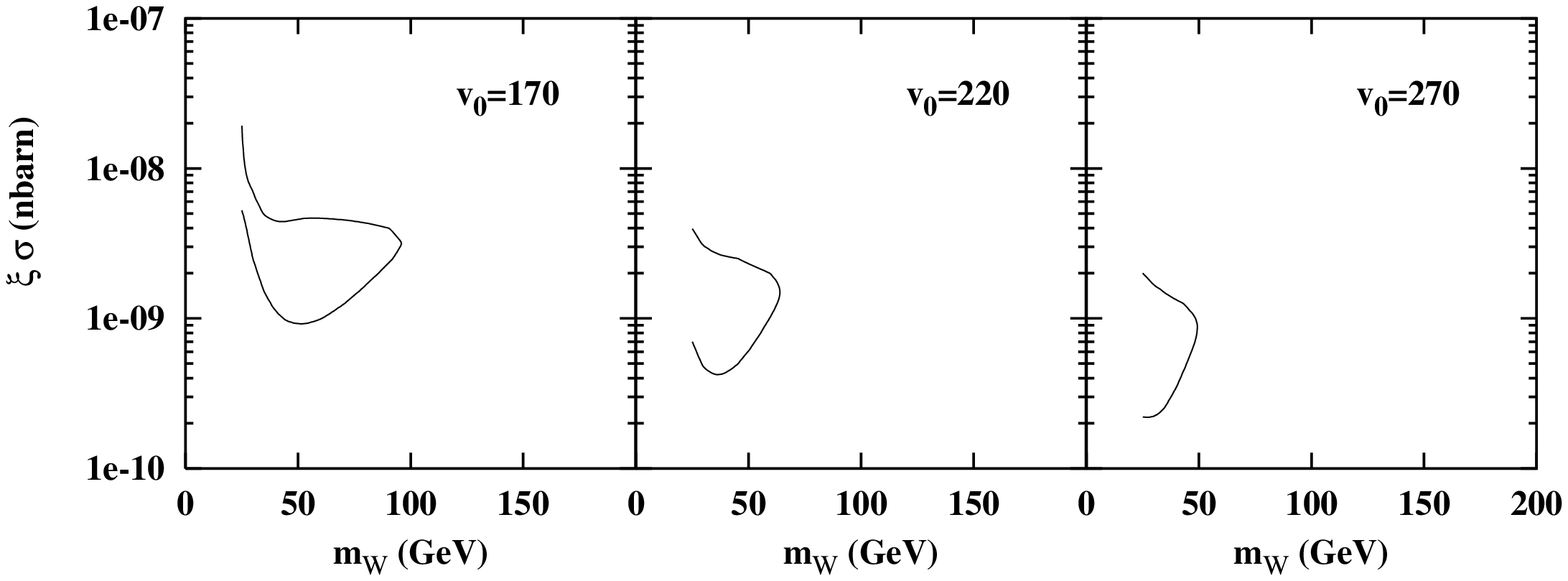}}
  \caption{
    The same as in Fig. \ref{fig:a0} for the velocity distribution of
    model C4. Only the case $\rho_0$=$\rho_0^{max}$
    is shown.
\label{fig:c4}}
\end{figure}

\begin{figure}\centering
  \rotatebox{270}{\includegraphics[bb=195 85 435 720,width=0.34\textwidth]{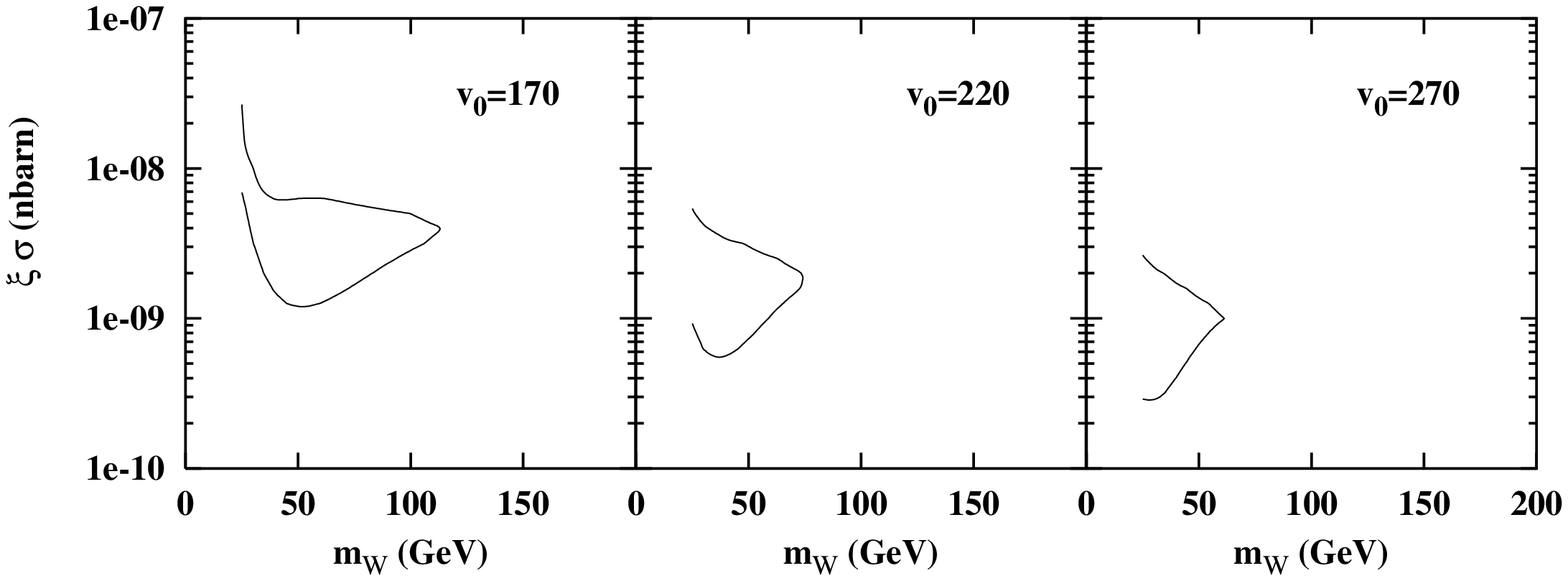}}
  \caption{
    The same as in Fig. \ref{fig:c4} including a co--rotation effect of the halo
    with $\eta=0.64$.
\label{fig:c4_corot}}
\end{figure}

\begin{figure}\centering
  \rotatebox{270}{\includegraphics[bb=195 85 435 720,width=0.34\textwidth]{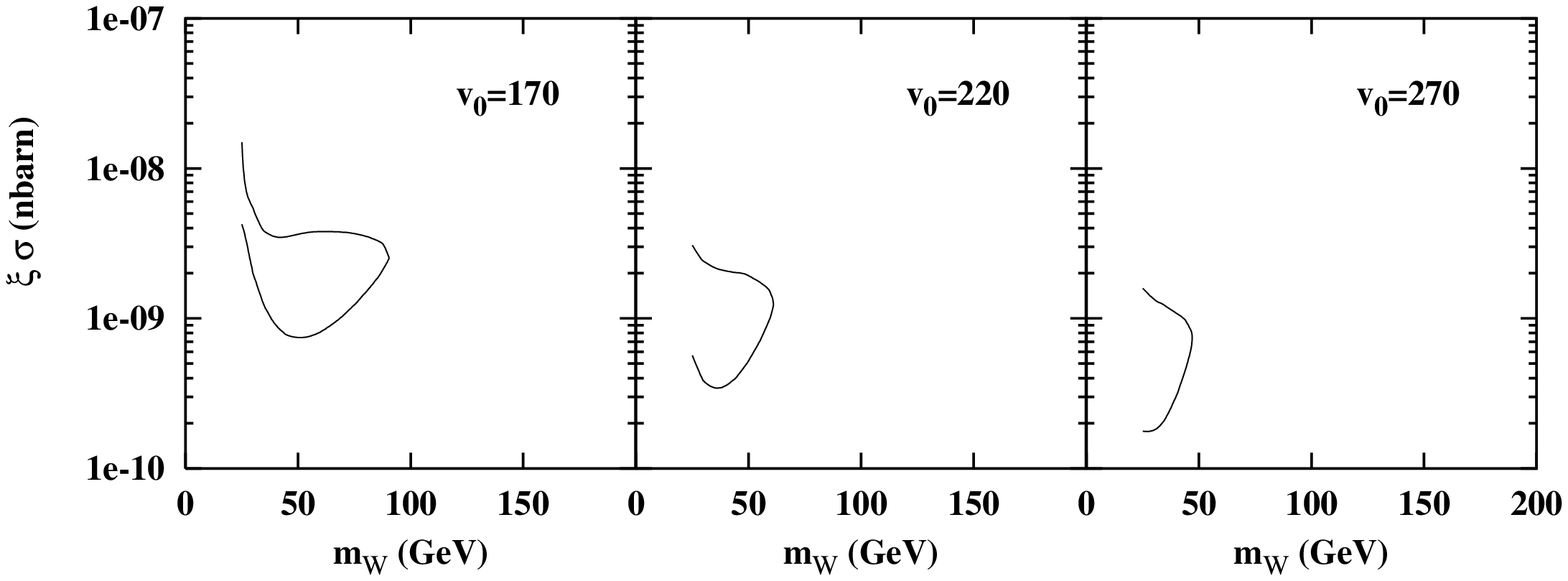}}
  \caption{
    The same as in Fig. \ref{fig:c4} including a counter--rotation effect of
    the halo with $\eta=0.36$.
\label{fig:c4_controrot}}
\end{figure}

\begin{figure}\centering
  \rotatebox{270}{\includegraphics[bb=195 85 435 720,width=0.34\textwidth]{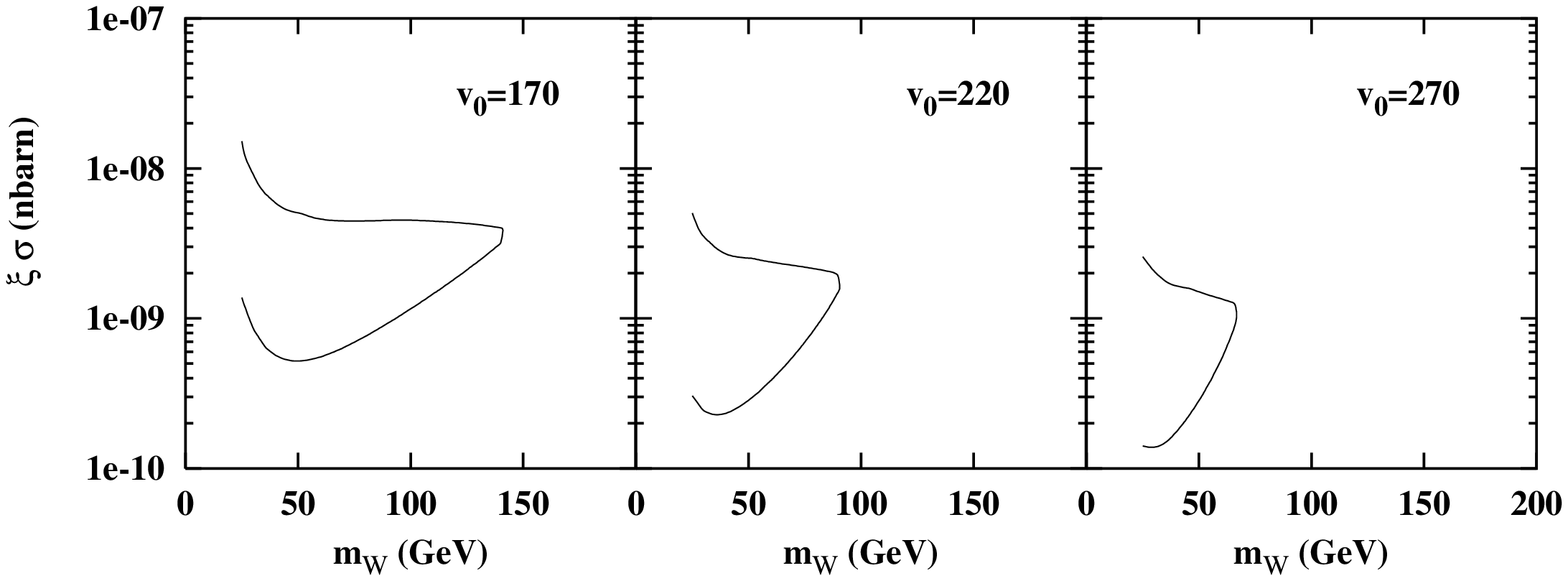}}
  \caption{
    The same as in Fig. \ref{fig:a0} for the velocity distribution of model D1.
    Only the case $\rho_0$=$\rho_0^{max}$ is shown.
\label{fig:d1}}
\end{figure}

\begin{figure}\centering
  \rotatebox{270}{\includegraphics[bb=195 85 435 720,width=0.34\textwidth]{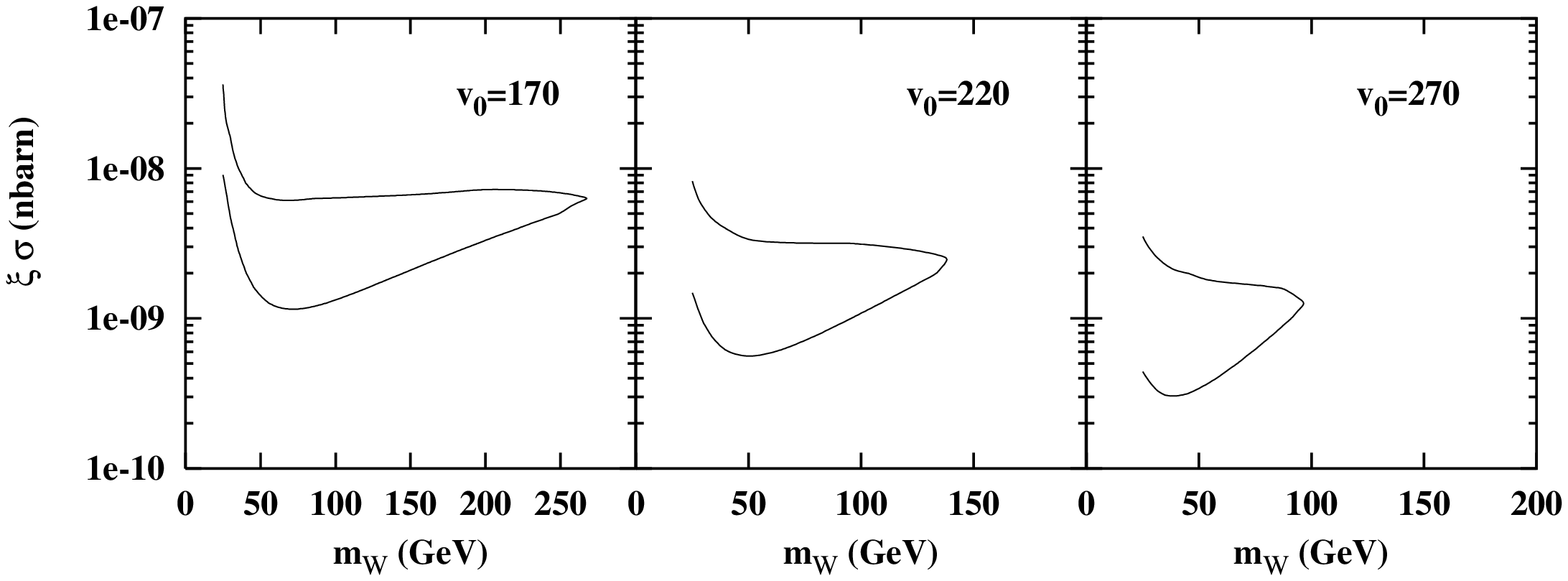}}
  \caption{
    The same as in Fig. \ref{fig:a0} for the velocity distribution of
    model D2. Only the case $\rho_0$=$\rho_0^{max}$
    is shown. The horizontal axis has been extended in the first panel.
\label{fig:d2}}
\end{figure}

\begin{figure}\centering
  \rotatebox{270}{\includegraphics[bb=195 85 435 720,width=0.34\textwidth]{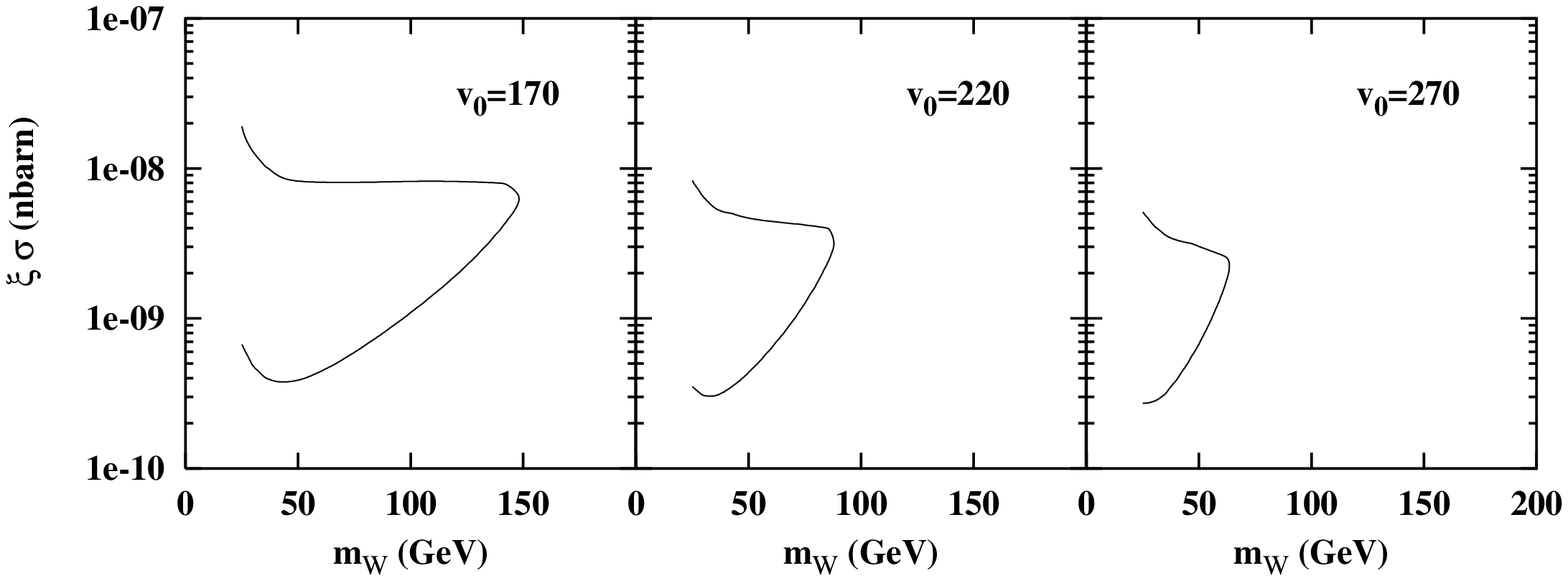}}
  \caption{
    The same as in Fig. \ref{fig:a0} for the velocity distribution of
    model D3. Only the case $\rho_0$=$\rho_0^{max}$
    is shown.
\label{fig:d3}}
\end{figure}

\begin{figure}\centering
  \rotatebox{270}{\includegraphics[bb=195 85 435 720,width=0.34\textwidth]{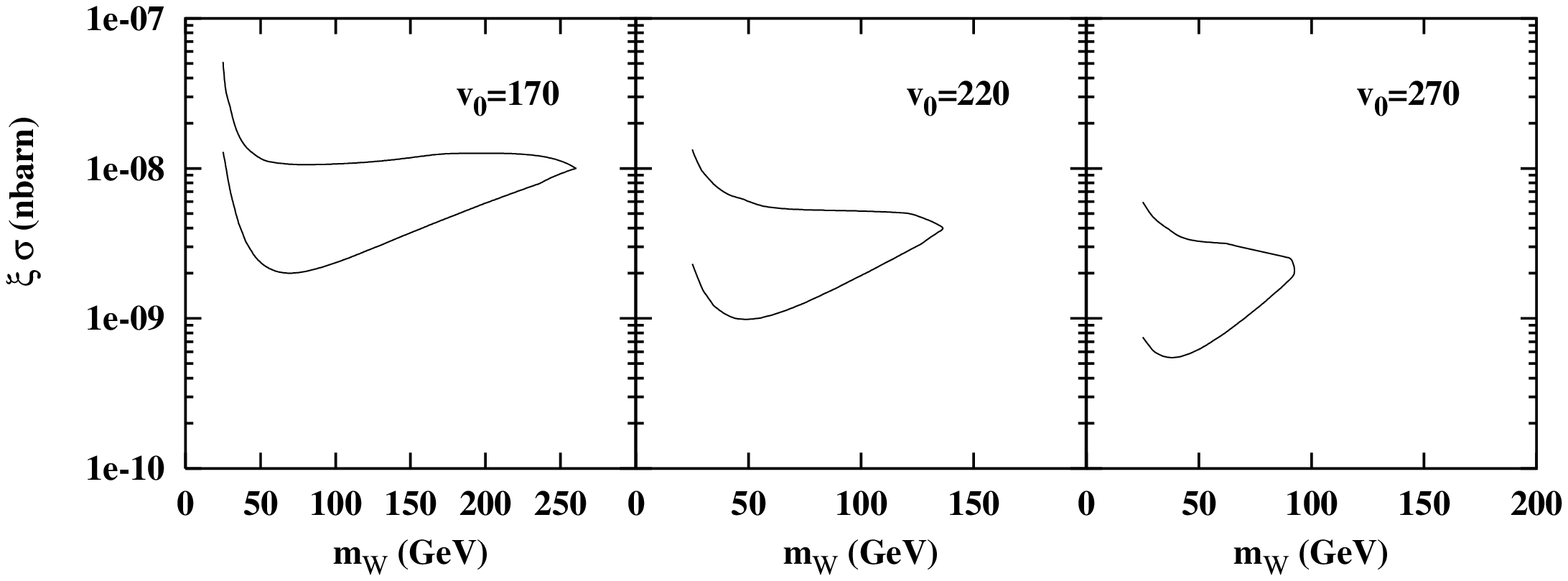}}
  \caption{
    The same as in Fig. \ref{fig:a0} for the velocity distribution of
    model D4. Only the case $\rho_0$=$\rho_0^{max}$
    is shown. The horizontal axis has been extended in the first panel.
\label{fig:d4}}
\end{figure}

\begin{figure}\centering
  \rotatebox{270}{\includegraphics[bb= 50 50 470 770, width=0.50\textwidth]{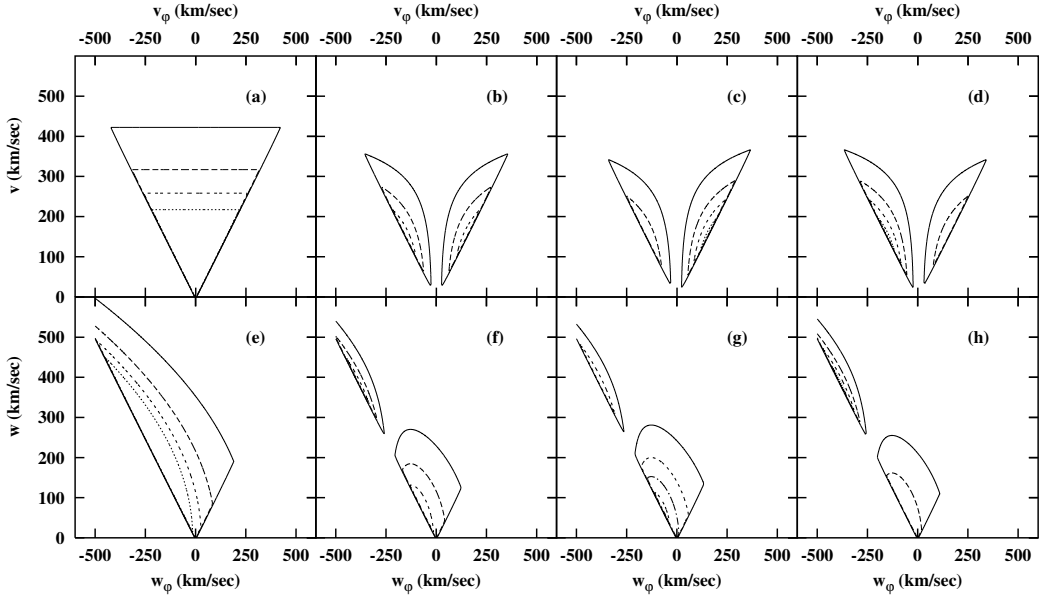}}
  \caption{Contour plots of the DF's for models B1 and C1 (see Table
    \ref{tab:models}). From left to right, the different panels refer to models
    B1 (panels (a) and (e)), C1 (panels (b) and (f)), co--rotating C1 (panels
    (c) and (g)), counter--rotating C1 (panels (d) and (h)).  Upper panels are
    plotted in the $v_{\phi}$--$v$ plane, defined in the reference frame of the
    Galaxy, while lower panels are shown in the $w_{\phi}$--$w$ plane, defined
    in the reference frame of the Earth. Solid lines, big dashes, small dashes
    and dots correspond to growing values of the DF (in arbitrary units).  The
    two disconnected closed contours which arise at different $w$ values in
    panels (f), (g) and (h) signal the superposition in the WIMP phase space of
    two components with well separated r.m.s. velocities.
\label{fig:test_c1}}
\end{figure}

\begin{figure}\centering
  \rotatebox{270}{\includegraphics[bb=195 85 435 720,width=0.34\textwidth]{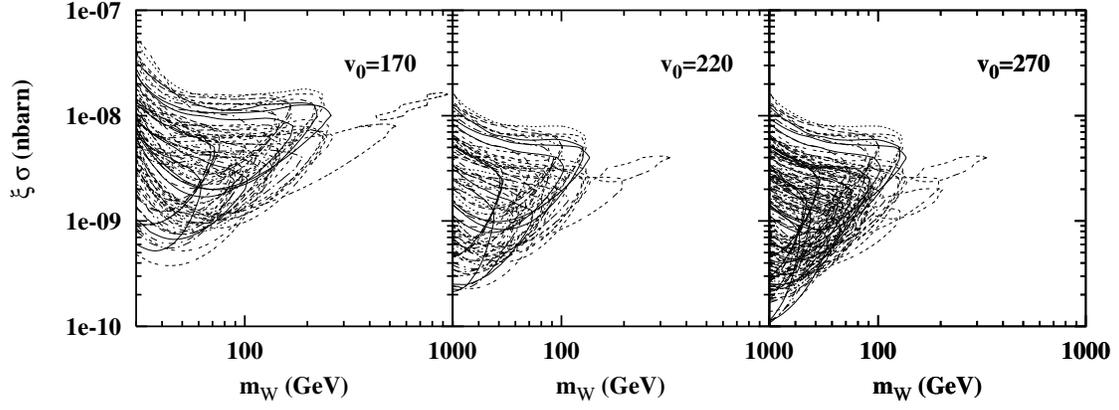}}
  \caption{Summary of the 3$\sigma$ annual--modulation regions in the plane $\xi
    \sigma_{\rm scalar}^{\rm (nucleon)}$ versus $m_{W}$, obtained by
    superimposing the results obtained with the velocity distributions of all
    the models described in Table \ref{tab:models}. For each of the models
    A1--7 and B1--7 two regions are plotted, which refer to the two extreme
    values $\rho_0^{min}$ and $\rho_0^{max}$ shown in Table \ref{tab:intervals}
    for the WIMP local density $\rho_0$. For models C1--4 and D1--4 only the
    regions which refer to $\rho_0=\rho_0^{max}$ are shown.
\label{fig:all}}
\end{figure}

\begin{figure}\centering
  \includegraphics[bb=195 85 435 720,width=0.34\textwidth]{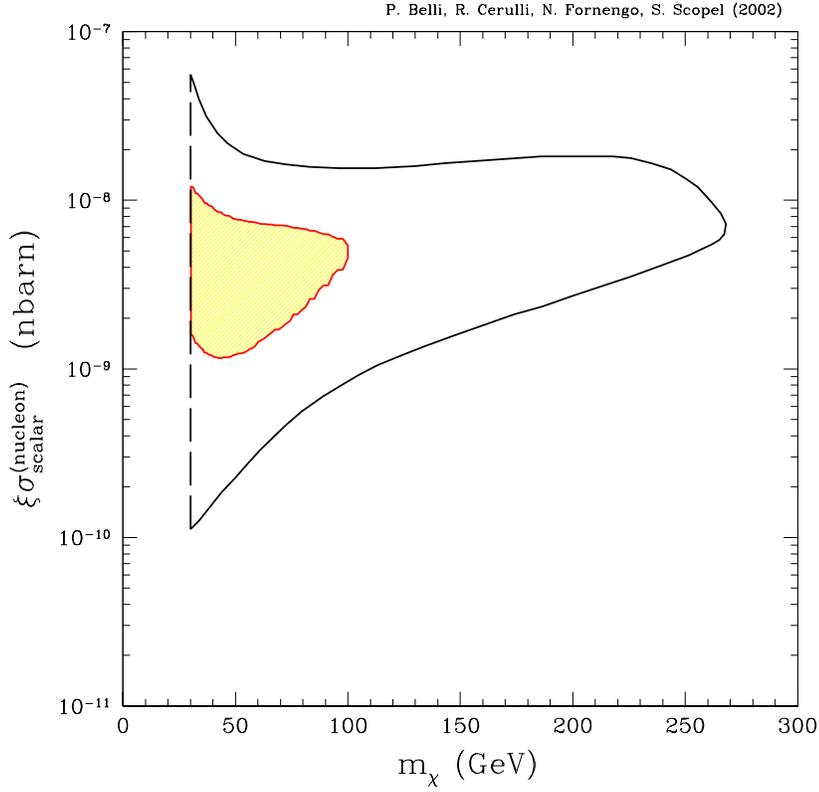}
  \caption{3$\sigma$ annual--modulation region in the plane $\xi
    \sigma_{\rm scalar}^{\rm (nucleon)}$ versus $m_{W}$, obtained by
    considering all the (non--rotating) galactic halo models discussed in this
    paper. The region is compared with the original annual modulation contour
    (shaded region) obtained in Ref. \cite{damalast} for an isothermal sphere
    model of the galactic halo with rotational velocity $v_0 = 220$ km s$^{-1}$
    and local dark matter density $\rho_0 = 0.3$ GeV cm$^{-3}$.
\label{fig:convolution}}
\end{figure}

\end{document}